# 6 Commons at the Intersection of Peer Production, Citizen Science, and Big Data: Galaxy Zoo

Michael J. Madison*

## I. Introduction

Policy analysis of scientific research, particularly in recent decades, has focused on tensions between norms of open science and knowledge sharing, on the one hand, and political and economic pressures to embed scientific research in market-based institutions based on proprietary claims to knowledge, such as modern patent law, on the other hand (Eisenberg 1989; Rai 1999; Reichman & Uhlir 2003). Twenty-first century technologies offer additional challenges and opportunities for science, grounded in the emergence of the Internet as a communications medium and in the explosion in the quantity of data available for study. If contests between norms of open science and the expectation that new knowledge should be propertized frame one (older, but still meaningful) debate about scientific research, then the emergence of so-called Big Data, often referred to more descriptively as data-intensive science (Hey et al. 2009), frames a second, related, and broader new debate. The new question is this: How should new scientific knowledge be governed? Do the historical poles—open science based on the norms of a scientific discipline, versus propertized knowledge grounded in the patent system—still offer the key alternatives? This chapter suggests that they do not. It offers a study of the organization and practices of scientific research in a contemporary astronomy project, Galaxy Zoo. The intuition explored below is that *commons*, rather than either the market or the social norms of science taken alone, offers a superior analytic framework for understanding the changing futures of science.

---

* Michael J. Madison is Professor of Law and Faculty Director, Innovation Practice Institute, University of Pittsburgh School of Law, Pittsburgh, Pennsylvania, USA.

Galaxy Zoo supplies a wildly successful model of what popular media refer to at times as "citizen science" and at other times and in other respects as peer production or "crowdsourcing." Academic researchers in astronomy in 2007 created a website that invited any and all comers to undertake the task of classifying approximately 900,000 galaxies, by looking at images downloaded from a recent sky survey. The classification exercise involved only a handful of relatively simple criteria and could be undertaken by nonexperts after a brief online tutorial. The sponsoring researchers expected to rely on the results as part of preparing traditional scientific papers. (In the main, that has been the case, with some exceptions, as discussed in more detail below.) The researchers had modest expectations at the start regarding the number of visitors to the site and regarding the length of time that completing the classification exercise would take. Not only were those expectations rapidly and vastly exceeded, but the number of volunteer classifiers and their enthusiasm led in short order to continuing the project both in depth and in breadth. The initial Galaxy Zoo enterprise has been extended multiple times, to include classification of additional astronomical data and also to facilitate additional and different types of social and scientific engagement by public volunteers. The sociotechnical "zoo" architecture that evolved in conjunction with the original Galaxy Zoo project has been refined and applied to additional and similarly structured scientific research projects, all of them collected since December 2009 under the umbrella name, the "Zooniverse."[1]

Galaxy Zoo and its success teach many things. Foremost among them, for present purposes, is the proposition that large-scale scientific research projects, particularly those involving extremely large sets of data, can be managed productively by threading a careful path between the idea that scientific research is (or should be) fully and completely open and the idea that large-scale knowledge-based enterprises are best governed via patent law. Galaxy Zoo relies on both openness and on property norms, and in that sense it serves as a useful case of commons.

"Commons," as the term and concept are used in this chapter, is an umbrella idea that refers to a broad array of possible institutional arrangements for sharing information and knowledge (that is, products and sources of human culture) and for sharing legal rights that might pertain to that information and knowledge. Accessible examples of such commons include patent pools; contemporary online, networked digital resources such as Wikipedia; modern analog knowledge production and distribution collectives such as news-reporting wire services; and older, conventional knowledge sharing institutions such as public lending libraries and universities.

---

[1] Zooniverse, https://www.zooniverse.org/. The citizen science/e-science projects housed at the Zooniverse as of this writing all are classified as relating to space, to the Earth's climate, to biology, to nature, and to the humanities. They include the Galaxy Zoo, the Moon Zoo, Solar Stormwatch, Planet Hunters, the Milky Way Project, Planet Four, Spacewarps, Old Weather, Cyclone Center, Ancient Lives, WhaleFM, Seafloor Explorer, Bat Detective, Notes from Nature, and Cell Slider. The original Galaxy Zoo was chosen for this case study based on the fact that its design and success were foundational to the rest.

Commons and commons governance refer to resource sharing and therefore to openness, but in present usage it refers to *structured* openness, with formal and informal institutional mechanisms in place to manage or govern that openness. Commons is governance, and commons is constructed, or built. In this sense commons should be distinguished from the unrestricted formal openness which defines the concept of the public domain in intellectual property law and which is sometimes attached to the term "commons" in casual, rhetorical, or political usage. I refer to commons as "constructed" because of the important sense in which commons are human institutions, often produced purposefully but sometimes emerging from or evolving out of historical happenstance. Below, I refer to the "knowledge commons research framework" as the analytic framework presented in Chapter 1 and in Madison, Frischmann, & Strandburg (2010).

The case study yields certain tentative conclusions and some hypotheses to be tested in future research. First, at a high conceptual level, there exists a dynamic relationship among scientific practice, forms of knowledge and knowledge structures, and social organization. This is hardly novel; it merely documents, again, what has been described by a diverse array of scholars, including Brown & Duguid (2000), Hilgartner & Brandt-Rauf (1994), Kuhn (1962), Mandeville (1996), Polanyi (1946), and most famously, Merton (1973). Second, at a more concrete level, relevant forms of social organization—including both the shape of astronomy and astrophysics disciplines and the character of their commons governance—are dependent on changing conceptual and material (technologically grounded) understandings of the data that scientists generate and use. Modern astronomy and astrophysics are, to significant degrees, large and complex exercises in information and knowledge governance, in addition to celestial observation. The growing computational character of these disciplines is in many respects akin to the computational character of contemporary geophysics. In light of that dynamism, none of these disciplines is itself static. What "is" science (or what "is" astronomy), what "is" relevant scientific research, and how individuals and groups bind themselves to those understandings are parts of the processes by which commons are constituted and by which commons constitutes those disciplines. I suggest, in other words, that institutional order and knowledge governance such as commons are mutually constitutive.[2] One exists largely because of the other. Finally, and despite the popular view that citizen science projects such as Galaxy Zoo primarily involve participation by undifferentiated "crowds" of volunteers, in practice the knowledge community that makes up the project succeeds in large part because it incorporates a defined social structure and a layered matrix of visions and purposes, from "this is astronomy" to "this is how the astronomical data should be classified," that both produce that social structure and are reinforced by it. More specific and detailed implications follow, below.

---

[2] For a related view based on a review of crowdsourced citizen science projects, see Mansell (2013).

The chapter is organized as follows. Part II describes Galaxy Zoo in broad outline and explains the motivations for this examination. Part III provides an overview of the knowledge commons concept detailed in Chapter I. Part IV applies the knowledge commons research framework to the Galaxy Zoo, both by supplying a brief narrative of its history and functioning and by breaking down its components in light of the detailed clusters of research inquiries suggested by the framework. Part IV also contrasts the Galaxy Zoo case with some hypothetical and actual institutional alternatives, in order to bolster the argument for evaluating Galaxy Zoo as knowledge commons. Part V offers implications and lessons, and Part VI concludes.

## II. Galaxy Zoo and Its Contexts

Humans have long looked to the stars to understand how they should look at each other and at their world. Humans likewise have long looked to each other to understand how they should look at the stars. That reciprocal relationship gave us the disciplines of astrology, astronomy, and astrophysics and ever greater understandings of literal and metaphorical influence and force. Along the way, and beginning with early astronomers, cultures of scientific inquiry and research emerged, with their own influence and force both on scientists themselves and on the institutions of science and related public policy. Comte (1830: 135–37) characterized astronomy as the first science and the foundation of all other sciences. Kuhn (1957) used the history of astronomy to frame a communitarian theory of the advancement of scientific knowledge. Traweek (1988) focused on an adjacent discipline, high-energy physics, to inaugurate the contemporary anthropological study of scientific disciplines as communities. The present chapter examines an astronomical research project partly because of the antiquity of the questions that the project addresses, partly because of the strong historical lineage of cultural questions of openness and access in communal terms, and partly because of the powerful and dramatic changes that these disciplines are now undergoing as a result of technological innovations described below.

Despite the cultural valence of earlier inquiries into the research communities of astronomers, physicists, and astrophysicists, the goal of this chapter is not to explore the anthropological or sociological character of the communities that constitute Galaxy Zoo. Rather, the goal here is to use this case to explore the dynamics of knowledge production, sharing, and dissemination in one particular scientific research community.

Galaxy Zoo began as a single solution to a pair of research problems. One of these was the domain of Kevin Schawinski, who in the early-2000s was a graduate student in astronomy at the University of Oxford. Schawinski was researching the evolution of elliptical galaxies (Clery 2011), that is, he was pursuing morphological analysis of galaxies, distinguishing elliptical from spiral-shaped galaxies. Galaxy morphology is closely linked to color. Most spiral galaxies have a distinct blue tinge, which is associated with the younger, hotter stars in their spiral arms; elliptical galaxies usually appear red, indicating the older ages of their stars and low levels of star formation. Blue ellipticals suggest the existence of gas reservoirs sufficient to support significant levels of star formation and are therefore of special interest to researchers (Schawinski et al. 2009). Schawinski aimed to examine a massive amount of digital astronomical data recently made available by the Sloan Digital Sky Survey (SDSS), a project of an international consortium of seven universities, other participating research institutions, several governments, and the Alfred P. Sloan Foundation (as principal funder). The SDSS had undertaken the largest comprehensive electronic map of the northern sky produced to date. Using a special purpose telescope on Apache Point, New Mexico, beginning in 2000 it imaged 10,000 square degrees of the sky, 70 million stars, and 50 million galaxies, resulting in approximately 15 trillion bytes of data, all of which were made publicly available as images to the research community (Margony 1999; Szalay et al. 2002). Schawinski planned to review and classify approximately 900,000 galaxies disclosed in the SDSS data. He tried, briefly, to do this himself, but he abandoned the effort because it was simply too time-consuming.

The second problem was the domain of another Oxford researcher, a postdoctoral fellow named Chris Lintott. Lintott was trying to understand spiral galaxies, also within the SDSS dataset. Whereas Schawinski was after blue ellipticals, Lintott was after red (that is, mostly dead) spirals (Lintott et al. 2008). (In each instance, the existence of these galaxies would suggest new research problems having to do with galaxy evolution and the birth and death of stars.) Schawinski's conversations with Lintott yielded the idea that the classification exercise that interested each of them could be outsourced, in a manner of speaking, to the public. Borrowing insights and some elementary technology from other, recent online scientific "crowdsourcing" efforts, notably Stardust@Home,[3] the first, public, Galaxy Zoo website (http://www.galaxyzoo.org) made the SDSS image data available online beginning in July 2007. The images were accompanied by a brief tutorial describing the classification dimensions that visitors were invited to learn and apply. (The phrase "Galaxy Zoo" evokes the idea of a zoo of galaxies—a somewhat unruly collection of "animals" with distinct appearances.) A handful of simple questions were asked, directed to morphological issues. Based on a brief online tutorial, users were asked: Is this an elliptical galaxy or a spiral galaxy, or something else? If it is a spiral galaxy, which way does it appear to be rotating? Related publicity (principally through the BBC) described the launch of the project and pointed visitors to the website (McGourty 2007).

The project was a tremendous success almost overnight, in several senses. In the first place, the classification problem that Galaxy Zoo was intended to solve was solved far more quickly and thoroughly than the organizers anticipated. "Within 24 hours of launch, the site was receiving 70,000 classifications per hour. More than 50 million classifications were received by the project during its first year, from almost 150,000 people."[4] Galaxy Zoo is now the world's largest database of galaxy shapes (Masters 2013).

---

[3] The Stardust@Home project began in 2006 and is located at http://stardustathome.ssl.berkeley.edu/. It engages ordinary citizens, who refer to themselves as "dusters," to examine movies of the universe to detect interstellar dust particles collected by the Stardust mission, launched by NASA in 1999 to explore the comet Wild 2.

In the second place, the model of citizen science data analysis that Galaxy Zoo introduced appears to have been accepted by the community of professional astronomers. The original Galaxy Zoo project is now complete, but it has been succeeded by follow-on astronomical research projects using closely related protocols: Galaxy Zoo 2, which asked participants to classify more finely a subset of 250,000 galaxies from the original SDSS Main Galaxy Sample, using a different and more detailed set of questions; Galaxy Zoo: Hubble, which asked participants to classify a different group of older and more distant galaxies using data derived from images obtained through the Hubble Space Telescope; and now Galaxy Zoo Quench, which offered volunteers the opportunity to both classify and analyze galaxy data. In the case of the original Galaxy Zoo, Galaxy Zoo 2, and Galaxy Zoo: Hubble, the zoo-produced data either has been incorporated into a continuing series of scientific research papers published in scholarly journals or is being prepared for publication. (Galaxy Zoo Quench is still in progress, as of this writing.) More than thirty peer-reviewed papers have followed from analysis of the original Galaxy Zoo data.[5]

In the third place, Galaxy Zoo has had unanticipated spillover benefits. Galaxy Zoo volunteers organized themselves into an online forum, creating and sustaining a community that is adjacent to and that in some respects overlaps with the community of professional astronomers. Through their forum, member "Zooites," that is, amateur Galaxy Zoo participants,[6] have made a number of important discoveries based on the original SDSS data shared via Galaxy Zoo, such as the object known as Hanny's Voorwerp and the so-called "Green Pea" galaxies (Fortson et al. 2009). Those discoveries have themselves been the bases for a number of scholarly papers. All of these Galaxy Zoo projects are now part of the larger cluster of citizen science projects described in the introduction to this chapter, known as the Zooniverse. Nearly 200,000 people are registered users of one or more Zooniverse projects, and more than 800,000 have participated in one way or another.[7]

The introduction to this chapter noted that exploring Galaxy Zoo as a commons case means observing an institution at the intersection of several distinct but sometimes overlapping phenomena: peer production or crowdsourcing; citizen science; and Big Data, or data-intensive science. Before turning to the knowledge commons framework and institutional analysis of Galaxy Zoo as commons governance, it is appropriate briefly to note the meaning and significance of these concepts in the Galaxy Zoo context.

---

[4] Galaxy Zoo, http://www.galaxyzoo.org/#/story, accessed June 12, 2013.
[5] A list of published papers appears at http://www.galaxyzoo.org/#/papers, accessed June 12, 2013.
[6] The term "Zooite" is, of course, highly informal. Its principal meaning captures volunteers who participate in the Galaxy Zoo forum, but its broader meaning includes all those who contribute to the classification project.
[7] The home page of the Zooniverse, https://www.zooniverse.org/, posts a tally of total participants.

Crowdsourcing is a popular and relatively recent term that denotes a product- or service-producing enterprise or exercise in which most of the relevant labor, if not all of the labor, is supplied from a so-called "crowd"—a large number of distributed individuals who are loosely coordinated, who offer their labor and/or expertise mostly as volunteers, and who are organized, if at all, mostly in nonhierarchical relationships. Benkler (2006) uses the phrase "peer production" to describe the phenomenon as follows. Some very large group of "peers," that is, individuals coordinated through a network rather than hierarchically through a firm, cooperatively and effectively produce some "thing," typically directed to knowledge, information, or culture, such as an online encyclopedia. The contents of Wikipedia are said to have been produced by crowdsourcing, for example, because of the very large number of volunteer Wikipedia contributors and editors and because the vast majority of them operate anonymously or pseudonymously and without formal supervision or discipline in a firm-based hierarchy.

It is important not to overstate the novelty of crowdsourcing or peer production, even if the Internet seems to make them much easier. The *Oxford English Dictionary* was produced in large part based on the contributions of thousands of volunteer lexicographers, who found and submitted slips of paper containing old literary references for unusual words (Ogilvie 2012; Winchester 2004). Nonetheless, the original Galaxy Zoo is in a sense an exemplary case of peer production or crowdsourcing. Thousands of anonymous and pseudonymous classifiers cooperated in producing an extraordinary database of morphologically classified galaxies.

The phrase "citizen science" captures a slightly different if sometimes related phenomenon and a different source of the significance of Galaxy Zoo. Citizen science describes the contributions of nonprofessionals to the work of professional scientists, in data collection and curation, analysis, or even production of scholarship based on that data (Cooper 2012a; Cooper 2012b; Reed et al. 2012). Citizen scientists may operate in large groups (in that sense, a large group of citizen scientists may crowdsource scientific data), in small groups, or even as solo contributors. Both halves of the phrase "citizen science" are significant. Individual nonprofessionals are engaged in *science* in some meaningful sense. And they are *citizens*, both in the sense that they are citizens of the world (that is, lay practitioners) rather than citizens of the scientific community and also in the sense that indirectly, at least, they subscribe to some flavor of the norms of citizenship that define that scientific community.

Although the phrase citizen science dates only from the early 1990s, the practice is, like crowdsourcing, relatively old. Examples come from both England and the United States. During the 1840s American sailors were charged by the head of the Depot of Charts and Instruments of the Navy Department with the systematic collection of weather data that were used to create wind and current charts (Cooper 2012a).[8] In England, during the 1830s, with the blessing of the British Admiralty, a scholar named William Whewell coordinated the collection of tidal information at more than 650 locations by volunteers who partook in the "great tidal experiment" (Cooper 2012b).[9] Charles Darwin relied on observations collected by a host of others, transmitted via post, for evidence consistent with his theory of evolution by natural selection (Browne 1996).

---

[8] One of the Zooniverse projects is titled "Old Weather" and involves the transcription of old ship logs.

Some might quibble with characterizing data collection or even data analysis as "science" and characterizing these historical examples or Galaxy Zoo (which clearly follows in their tradition) as science. Stardust@Home, one of the modern models for Galaxy Zoo, is "science" in the same sense. Individual Stardust@Home citizen "dusters" and Galaxy Zoo citizen "Zooites" are credible citizen scientists because they actively apply their own critical faculties in selecting and/or analyzing data.[10] As discussed below, that identity appears to be an important factor in the success of Galaxy Zoo. From the perspective of citizen volunteers, SETI@Home, a longer-standing project through which individuals allow spare computing cycles in their Internet-connected computers to be harnessed as part of a larger-scale "distributed" computing project that analyzes radio signal data to detect signals that might indicate the existence of extraterrestrial intelligence, likely is not science. The distinction is the nature of the individual agency involved. SETI@Home is a scientific research project, and its data is peer produced (or crowdsourced), but volunteer participants are not necessarily citizen scientists. They have merely loaned their technology to the actual researchers and become passive participants in a computing exercise.

Big Data is the third significant domain in which Galaxy Zoo is situated. Contemporary commentators now often use the more prosaic and less media-friendly phrase "data-intensive science" to refer to what is simultaneously field, problem, and opportunity: how to make productive use of the massive quantities of digital data now made available by widely distributed sensor networks and highly sensitive observing instruments coupled with massive data storage facilities (Manyika et al. 2011). Exploring, classifying, searching, visualizing, and otherwise using this information—which is quantified in terms of petabytes—now constitutes a healthy research agenda in itself, because doing all of these things requires and will continue to require a host of new technological tools. Both *Nature* and *Science* have devoted special issues in recent years to the nuances of the challenges of Big Data, testifying to the scale and significance of this development (Nature 2011; Science 2011). The National Research Council of the National Academies of Science recently released a report synthesizing contemporary thinking across the scientific and policy domains implicated by data-intensive science (National Research Council 2012). Much of the report is taken up by surveys of the scientific disciplines that have begun to tackle Big Data challenges, notably geophysics and astronomy. It also directs attention to underlying governance problems in light of the fact that much of this data is being generated and distributed in open networked environments funded by many different institutions and researchers. How will this data be managed?

---

[9] William Whewell is also credited with being the first to publicly associate the term "scientist" with a person who practiced science, although the term did not achieve widespread acceptance for several decades (Whewell 1840: cxiii). The terms "philosopher" and "natural philosopher" were no longer sufficient to distinguish among practitioners of different scientific disciplines.

[10] This summary intentionally excludes the vast emerging infrastructure of online networked technical resources that bring observational opportunities to astronomical amateurs and professionals like by bringing high-resolution imaging to the desktop, including Google Sky (http://www.google.com/sky/) and the Microsoft WorldWide Telescope (http://www.worldwidetelescope.org/).

Galaxy Zoo is an obvious candidate for discussion as an exemplar of a Big Data project. The SDSS dataset that formed the foundation of Galaxy Zoo constituted more than 10 terabytes of data. It is only one of several data releases provided by SDSS. Future SDSS surveys will produce even more data; technology improvements mean that astronomical data is likely to be created at the rate of multiple terabytes *per night* (Goodman & Wong 2009; Lawrence 2009; Škoda 2007). It is not only the case that new technologies and organizational models must be created and/or adapted to analyze that data but also that those technologies and models must be conceptualized. Peer production and citizen science are two possible conceptual sources that inspire and are inspired by Galaxy Zoo itself, but the challenges of data-intensive science are greater than Galaxy Zoo alone.

Making Galaxy Zoo the primary subject of the present chapter is prompted by interest in the sources and implications of each of these three perspectives. The type of data considered by Galaxy Zoo might lead observers to conclude that classification and aggregation of that data—the governance challenge—should be either a case of pure open science (as a traditional research scientist might suppose), or a case of market-based transfers of intellectual property (IP) rights (as a conventional intellectual property lawyer might suppose). In other words, Galaxy Zoo might have been conceived as a large version of a conventional research protocol, or as a marketplace for thousands of small-scale IP transactions. Each of the three perspectives highlighted above suggests that Galaxy Zoo is, to a significant degree, both more than and different from either of these things. Each of those perspectives is, in itself, incomplete. Treating Galaxy Zoo solely as a case of peer production or as a case of citizen science captures a great deal of the labor dynamics of the project but fails to capture fully its normative dimensions as scientific research. Treating it as a case of Big Data goes further in explaining those normative dimensions but potentially undervalues the significance of the contributions of the citizen scientists. Looking for transactions undervalues the seamlessness of the work in practice and the collaborative character of the relationships among the astronomical experts and the volunteers. What follows in the remainder of this chapter is largely an effort to synthesize the insights of each of those perspectives, using the broader rubric of the knowledge commons research framework. The hypothesis here is not that Galaxy Zoo "is" a commons, but instead that the commons framework offers a particularly useful way to understand how and why Galaxy Zoo functions as it does.

### III. The Knowledge Commons Framework

This part reviews the knowledge commons research framework, outlined in detail in Chapter 1 of this volume, for the sake of the internal completeness of this chapter.

Readers who are acquainted with the purposes and details of the framework may skip ahead to Part IV.

The knowledge commons framework builds on a series of related intuitions (Madison, Frischmann, & Strandburg 2010). Commons governance means knowledge and information management characterized by domains of managed openness and sharing of relevant resources, and the first intuition is that commons governance is in broad use in day to day practice in a variety of domains and across a variety of scales. Documenting evidence to justify that intuition is the first goal of the framework. The second intuition is that such structured openness in the management of both natural and cultural resources is likely to lead to socially beneficial and/or socially productive outcomes. Salient among the class of cases where commons governance is successful and sustainable are contexts where social interest in positive spillovers from bilateral, market transactions is high. Commons may sustain the production of spillovers when the market otherwise may not. Testing that intuition by applying the framework in a series of case studies is the second goal of the framework. The final intuition is that a standard framework for identifying and assessing commons across a variety of domains can support the development of more sophisticated tools for realizing the potential for commons solutions in new institutional settings and for distinguishing commons solutions from other solutions in settings where some other approach, such as an approach grounded in proprietary rights, might be preferred. Applying the knowledge commons research framework is an exercise in analyzing colloquial commons institutions, such as "scientific research" taken in the aggregate (Merges 1996), in a nuanced way via comparative institutional analysis.

Examining constructed commons in the cultural context builds on the Institutional Analysis and Development (IAD) framework pioneered by Elinor Ostrom and her colleagues (Ostrom 1990; Ostrom 2011), but it adds some important modifications. The IAD framework has been used principally to structure analysis of solutions to collective action problems in natural resource contexts (so-called action arenas, or action situations) such as forests, fisheries, and irrigation systems. IAD analysis is premised largely on choice-processing, goal-oriented behavior by self-interested individuals. (Individual agents may fully informed and rational or operating under conditions of bounded rationality.) It looks to explain sustainable collective action that produces measurable, productive results. The insight from applying the IAD framework to a large number of governance institutions and resources is that commons solutions can be as robust as market-oriented solutions to classic "tragedy of the commons" overconsumption dilemmas involving depletable natural resources. Shared governance can lead to sustainable fisheries and forests and to regular supplies of usable water.

The knowledge commons framework differs in certain key respects. It does not assume the agency of rational, choice-selecting, self-interested individuals. It accepts the role of historical contingency and of inward-directed agents in the evolution of collective or commons institutions. At the front end of the analysis, it also requires understanding the contingency of the underlying resources themselves. Natural resource commons largely take the existence of their resources for granted: fish, trees, water, and the like. Knowledge commons identify resource design and creation as variables to be analyzed. As cultural resources (that is, as forms of knowledge and information), patents, copyrights, and underlying inventions, creations, and data are shaped by a variety of institutional forces rather than by nature.[11] Critically, the knowledge commons framework does not assume that the relevant resources are rival and depletable. The knowledge commons framework generally assumes precisely the contrary: that intangible information and knowledge resources are nonrival, nonexcludable public goods. The dilemma to be solved is not primarily a classic "tragic commons" overconsumption problem. Instead, it is more likely (in part) an underproduction problem and (in part) a coordination problem. In the absence of a governance mechanism to moderate consumption, producers of resources will fail to invest in creating new goods or in preserving them, either on their own or in combination with others, because of uncertainty regarding their ability (either individually or collectively) to earn returns that justify the investment.

Against that background, the knowledge commons framework proposes to undertake comparative institutional analysis by evaluating cases of commons resources via a series of questions, or clusters, to be applied in each instance. Several of these are borrowed or adapted from Ostrom's IAD framework. Some are developed specifically for the knowledge commons context. The full list is described here, and an abbreviated version is applied to Galaxy Zoo in Part IV.

The initial question is whether the relevant resource or case is characterized from the outset by patent rights or other proprietary rights, as in the case of a patent pool, or by a legal regime of formal or informal openness, as in the case of public domain data or information collected in a government archive. A particular regime might involve sharing data and information, or sharing rights in information, or sharing both. The character of the commons solution might involve coordinating holders of different IP interests or holders of different public domain knowledge resources, for example.

Answering that question sets a default baseline against which a commons governance regime is constructed. Within that regime, one next asks definitional questions. What are the relevant resources, what are the relationships among these resources, the baseline, and any relevant legal regime (for example, what a scientist considers to be an invention, what patent law considers to be an invention, and the boundaries of the patent itself are three related but distinct things), and what are the boundaries and constitution (membership) of the community or communities that manage access and use of those resources? How is membership acquired (this may be informal, formal, or a blend of the two), and how is membership governed? What is good behavior within the group, what is bad behavior, who polices that boundary, and how?

---

[11] Institutional forces and technologies bear on the constitution of natural resources as well, to be sure. But the biophysical dimensions of resources in natural resource commons analysis are both materially and conceptually different from the metaphysical dimensions of knowledge and information "things" in knowledge and information commons (Madison 2005).

Next are questions concerning explicit and implicit goals and objectives of commons governance, if there are any. Is there a particular resource development or management dilemma that commons governance is intended to address, and what commons strategies are used to address that dilemma?

How "open" are the knowledge and information resources and the community of participants that create, use, and manage them? Madison, Frischmann, & Strandburg (2010) argue that commons governance regimes involve significance measures of resource and community sharing and openness. Their details, however, should be specified, along with their contributions to the effectiveness of commons. Some commons and commons resources have precise and fixed definitions of both resources and community membership. Or, either resources or membership or both may be more fluid, with boundaries defined by flexible standards rather than by rules.

A large and critical cluster of questions concerns the dynamics of commons governance, or what Ostrom refers to as the "rules-in-use" of commons: the interactions of commons participants and resources. Included in this cluster of questions are (1) stories of the origin and history of commons; (2) formal and informal (norm-based) rules and practices regarding distribution of commons resources among participants, including rules for appropriation and replenishment of commons resources; (3) the institutional setting(s), including the character of the regime's possibly being "nested" in larger scale institutions and being dependent on other, adjacent institutions; (4) relevant legal regimes, including but not limited to intellectual property law; (5) the structure of interactions between commons resources and participants and institutions adjacent to and outside the regime; and (6) dispute resolution and other disciplinary mechanisms by which commons rules, norms, and participants are policed.

At this point it becomes possible to identify and assess outcomes. In Ostrom's IAD framework, outcomes are typically assessed in terms of the resources themselves. Has a fishery been managed in a way that sustains fish stocks over time? Do commons participants, such as the members of a fishing community, earn returns in the commons context that match or exceed returns from participation in an alternative governance context? In knowledge commons, resource-based outcome measures may be difficult to identify and assess. Sustaining the resources and their uses, individually or in combination, may be the point. In a patent pool, pooled resources may constitute components of larger, complex products that could not be produced but for the pooling arrangement that reduces transactions costs among participants. Outcomes take different forms. It may be the case that patterns of participant interaction constitute relevant outcomes as well as relevant inputs. Agency, in a manner of speaking, may be less important than identity; the group and its participants, in a particular institutional setting, may be ends as well as means. Levels of interaction and combination matter. Participant interaction in the context of a shared resource pool or group may give rise to (or preserve, or modify) an industrial field or a technical discipline. In that specific case, such spillovers may be treated as relevant outcomes. One might characterize these phenomena *ex ante* as different "types" of commons (Benkler, this volume, ch. 3). The knowledge commons framework deems them all fair game for comparative inquiry.

Having identified relevant outcomes, it becomes possible to look back at the problems that defined commons governance in the first place. Has the regime in fact solved those problems, and if not, then what gaps remain? How do the outcomes produced by commons governance differ from outcomes that might have been available if alternative governance had been employed? On the other side of the assessment ledger, has commons governance created costs or risks that should give policy makers and/or institution designers pause? Costs of administration might be needlessly high; costs of participation might be high; and commons governance might offer a risk of negative spillovers that offsets the initial instinct that commons produce positive spillovers. A collection of industrial firms that pool related patents in order to produce complex products may produce those products but may also engage in anticompetitive, collusive behavior. Commons governance may facilitate innovation; it may also facilitate stagnation, or worse.

## IV. Galaxy Zoo as Knowledge Commons

In this part, the knowledge commons framework is applied to Galaxy Zoo. The results are derived primarily from examining publicly accessible online materials that describe and implement the project (http://www.galaxyzoo.org), which are extraordinarily rich and detailed. The information is supplemented by personal interviews that I conducted with the initiator of the project, Kevin Schawinski; with Brooke Simmons, a researcher at the University of Oxford and previously (with Kevin Schawinski) at Yale University; and with Kevin Schawinski's initial collaborator at the University of Oxford and the coordinator of the emerging Zooniverse, Chris Lintott. I have also relied on a number of research papers and conference posters reporting on studies of and by Galaxy Zoo participants. The results are reported below initially according to the clusters of questions framed by the knowledge commons research framework, although as will become clear, not all of those questions are useful or applicable here. Following that discussion, I consider the strengths and weaknesses of Galaxy Zoo as commons in light of possible alternative institutional frameworks. In that setting I briefly discuss a different Big Data astrophysics scientific research project, the Nearby Supernova Factory, that appears to be governed not as commons but instead as something approximating a traditional firm.

### A. GALAXY ZOO: APPLICATION OF THE FRAMEWORK

Applying the knowledge commons framework begins with some supplemental detail about the origins and outcomes of Galaxy Zoo.

Galaxy Zoo emerged from efforts to answer a research problem. The broad challenge, and the focus of the discussion below, was morphologically classifying roughly 900,000 known galaxies revealed in images collected in the Sloan Digital Sky Survey. The narrower challenges consisted of identifying blue elliptical galaxies, which are relatively rare, and red spiral galaxies. The results would (and now do) form the backbone of a database about galaxy types to help astronomers understand the evolutionary dynamics of galaxies. Schawinski, then a University of Oxford graduate student, had attempted initially to do the blue elliptical classification work himself. In a week's time, working full-time at the task, he classified 50,000 galaxies, separating spirals from ellipticals. He realized quickly that this approach was not likely to be time-effective, but his classification rate (50,000 galaxies in a week) became known within the Galaxy Zoo community as a "Kevin week," a unit of analysis (Adams 2012).

Schawinski and an Oxford postdoc also interested in galaxy morphology, Chris Lintott, determined that the classification task could not be undertaken accurately by computer technology acting alone. The images were too indistinct for image- and pattern-recognition software of the time to be trusted (Lintott et al. 2008). Astronomers had been looking for solutions in various computation-based methods: artificial neural networks, computational algorithms, and model-based morphologies coded into software. Instead, Lintott and Schawinski and a third Oxford researcher, Kate Land (working with a Slovenian astronomer, Anže Slosar), who was interested in the "handedness" of spiral galaxies (do they seem to spin to the left or to the right?), took their inspiration from the Stardust@Home citizen science project. Via a website tutorial and an invitation distributed initially in July 2007 with a press release picked up by British television (Lintott was a copresenter on the BBC's Sky at Night program) and the online BBC News and repeated via other online media, they invited public volunteers to answer a brief set of questions about galaxy images displayed on a public website (Fortson et al. 2009).[12] The central questions were: Is the galaxy elliptical or spiral? Does it appear to be something other than a galaxy, such as a star? Does the image appear to contain galaxies that are merging? And if the galaxy is spiral, is it spinning counterclockwise or clockwise, or does the rotation appear to be "edge on" (facing the viewer)? Volunteers were asked to register an account (a username or pseudonym was accepted) and to run through a trial set of fifteen sample galaxies. If they correctly classified eight or more, then they were permitted to move ahead into classifying further images.[13]

The scale of the response was unexpected and overwhelming. The researchers anticipated five to ten classifications per galaxy over a period of three years, so that the entire classification exercise would be completed in five years' time. Instead, within forty-eight hours of the site's going live, Galaxy Zoo was receiving more than 50,000 classifications per hour. The three-year target of 30,000 registered users was reached before the end of the first week. By the end of the first year, roughly 150,000 registered volunteers had submitted more than 50 million classifications. Because the servers storing the galaxy images did not withdraw an image after it had been classified previously, each galaxy had been classified more than 30 times. The original Galaxy Zoo project closed in February 2009. The accumulated data was publicly released and remains available at http://data.galaxyzoo.org/ (Lintott et al. 2010).[14] Papers analyzing the data are collected at https://www.zooniverse.org/publications.

---

[12] A contemporaneous blog post by Chris Lintott captures the history in more minute detail (Lintott 2007). A professional design team was retained to assemble the original Galaxy Zoo website.
[13] The full original Galaxy Zoo site is still online in archive form at http://zoo1.galaxyzoo.org.

Unexpectedly but noteworthy for present purposes, the explosion of volunteer interest in the project was accompanied by a demand by volunteers for answers to thousands of both scientific and technical questions. At least one celebrity with a noted interest in astronomy, Brian May, former lead guitarist of the band Queen and now a PhD astrophysicist, publicly promoted the Galaxy Zoo project, a fact that undoubtedly help stir public interest. But the sheer volume and intensity of the response suggests that Galaxy Zoo tapped into a huge, latent demand for amateur scientific—particularly astronomical—exploration. The Galaxy Zoo team launched a blog in which they discussed their work with the galaxy classification data. An online forum (still at http://www.galaxyzooforum.org) was added quickly—within a month of the launch of Galaxy Zoo—where volunteers could post interesting images and questions and talk among themselves as well as with the team leaders. Importantly, the images on the Galaxy Zoo site were served directly by the SDSS archive (more on the SDSS, below), and each image was accompanied by a link, via a disclosed galaxy reference number, to the SDSS Object Explorer for that image (currently located at http://skyserver.sdss.org/public/en/tools/explore/obj.asp). The Object Explorer allows users to explore objects in that image in greater details. The Galaxy Zoo exercise asked volunteers one relatively limited set of questions, but the website served volunteers an opportunity to explore the sky with very sophisticated tools.[15]

The Galaxy Zoo forum has remained an integral part of Galaxy Zoo and its successors (Galaxy Zoo 2, Galaxy Zoo: Hubble, and Galaxy Zoo Quench). Users of the forum, which has evolved into a longer list of moderated message boards, refer to themselves as "Zooites." From the beginning, moderation has been handled largely by volunteer Zooites who were appointed, informally, by Lintott. In turn, Zooites refer to the professional astronomers who lead the Galaxy Zoo team (at least partially in jest) as "zookeepers."[16] Among the successes of the Zooite community are the discovery of and published scholarship regarding additional "irregular" astronomical objects not targeted for initial analysis by the team (the object known as "Hanny's Voorwerp," discovered by a Dutch Zooite volunteer named Hanny Van Arkel, and a group of so-called "Green Pea" galaxies) (Clery 2011), and the fact that some of the more active volunteers have been inspired to move from amateur astronomy to acquisition of professional, academic training in the discipline (Adams 2012).

---

[14] Each volunteer's classifications are treated as "votes," and the Galaxy Zoo team developed software that converted these votes into final classifications. The Galaxy Zoo data release includes full detail regarding the number and distribution of classification "votes" for each image.

[15] A related SDSS tool, the SDSS Viewer, at http://ncastro.org/Contrib/SDSS/SdssViewer.htm, created additional opportunities for what Cohen (2012) might refer to as "play." Much like a contemporary online digital music service, using the Viewer SDSS users or groups of SDSS users could explore attributes of SDSS objects and create playlists of their favorite images. Those playlists are public, along with the image data itself. Using the link above, explore "GZ Take My Breath Away," "GZ Best Spirals Short List," and "GZ Best Spirals Full" playlists.

[16] A list of current and former zookeepers is posted at http://hubble.galaxyzoo.org/team.

I now turn to the clusters of commons-related questions suggested by the knowledge commons framework.

1. The Character of Commons Resources and of the Community

What are the relevant governable resources relevant to Galaxy Zoo? The resources here are of several kinds. The original resources are the underlying SDSS image data; Galaxy Zoo volunteers were presented with images of galaxies from this data archive. The next set of resources is the individual time and expertise of each Galaxy Zoo team member and volunteer, which is translated into some number of galaxy classifications. The third and final set of resources is the aggregated Galaxy Zoo data and the scholarly papers that resulted from it. Each set of resources should be characterized and analyzed in turn.

The SDSS image data deserve the greatest exploration and explanation. However one might characterize the innovation and impact of Galaxy Zoo, there is no doubt that the Sloan Digital Sky Survey is, itself, a transformative resource for astronomers (Finkbeiner 2010). The Sloan Digital Sky Survey is a seven-university cooperative enterprise that will result in the most precise, high-resolution electronic map of the northern sky ever realized. It has been analogized to a cosmological Human Genome Project (Hey 2013). The technical specifications alone suggest its massive scope. The SDSS was launched in 1998 and relies on data gathered from the Apache Point Observatory, a dedicated, 2.5-meter telescope on Apache Point, New Mexico operated by the Astrophysical Research Consortium that is equipped with a 120-megapixel camera. The camera can image 1.5 square degrees of sky at a time (about eight times the area of the full moon), and a pair of related spectrographs fed by optical fibers can measure spectra of (and therefore distances to) more than 600 galaxies and quasars in a single observation (York et al. 2000). A custom-designed set of software pipelines manages the enormous data flow from the telescope. The SDSS completed its first phase of operations, SDSS-I, in June 2005, and its second phase, SDSS-II, in 2008.

Over the course of five years, SDSS-I imaged more than 8,000 square degrees of the sky, detecting nearly 200 million celestial objects, and it measured spectra of more than 675,000 galaxies, 90,000 quasars, and 185,000 stars. Annual data releases beginning in 2005 and extensions of the SDSS phase of operations mean that the volume of data disclosed to the public, initially measured in terabytes, is now measured in petabytes. (The first and second SDSS programs can be accessed at http://www.sdss.org; the continuing program, SDSS-III, is accessible at http://www.sdss3.org. The SDSS-I and SDSS-II data releases form what is referred to as the "Main Galaxy Sample" and were the foundations for the original Galaxy Zoo.) As one scholar put it as the survey got underway:

> The imaging survey of 10,000 square degrees of the sky, 70 million stars, and 50 million galaxies will result in approximately 15 trillion bytes of data, an amount that will rival in quantity the total information content of the Library of Congress. Even amateur astronomers will be able to use the data to search for rare celestial objects, to study particular regions of the sky, or to search for unexpected similarities or differences in collected data (Fuchs 2001: 51).

Funding for the creation and distribution of the SDSS archive has been provided by the Alfred P. Sloan Foundation, the "Participating Institutions," the National Aeronautics and Space Administration, the National Science Foundation, the U.S. Department of Energy, the Japanese Monbukagakusho, and the Max Planck Society. The SDSS is managed by the Astrophysical Research Consortium (ARC) for the Participating Institutions. The Participating Institutions are the University of Chicago, Fermilab, the Institute for Advanced Study, the Japan Participation Group, the Johns Hopkins University, Los Alamos National Laboratory, the Max Planck Institute for Astronomy (MPIA), the Max Planck Institute for Astrophysics (MPA), New Mexico State University, University of Pittsburgh, Princeton University, the United States Naval Observatory, and the University of Washington.[17]

Making SDSS data accessible and usable to astronomers has required more than simply dumping the data onto servers linked to the Internet. Critical to use of SDSS image data generally and to the operation of Galaxy Zoo in particular was the development of the Sky Server, a collection of databases and associated websites and interfaces that offer SDSS data to the public. Production and management of the Sky Server (currently located at http://cas.sdss.org/dr7/en/ and hosting SDSS Data Release 7) is coordinated principally among Johns Hopkins University (with respect to the underlying data archive) and Microsoft Research (with respect to the Sky Server technology itself). When the Galaxy Zoo website went live, it was hosted on Johns Hopkins University servers (later, because of the volume of traffic, it was moved to a commercial vendor). As noted above, the images themselves were served to Galaxy Zoo volunteers, via the Galaxy Zoo website, directly from the Sky Server.

The abundant technical and organizational detail that accompanies discussion of the SDSS data are important in part to demonstrate the complex and layered character of the source knowledge resources that lie at the center of Galaxy Zoo. A massive and sensitive astronomical instrument obtained a daunting amount of data, which was processed, sorted, and organized into particular images stored in a networked data archive and then made available for public viewing and analysis via yet another sophisticated collection of information technologies. In one sense, these images and the collection of galaxies they represent are wholly manufactured by experts using sophisticated technology, and the scope of the resource pool is limited by the scope and pace of the experts' time and manufacturing capabilities. At some formal or technical level, moreover, each image and the observable characteristics regarding the objects in that image are unique, and they are necessarily unique because they relate to an underlying physical reality.

---

[17] A thorough yet accessible summary of the SDSS is available at http://cas.sdss.org/dr7/en/sdss/. The Sky Survey is part of an international consortium of publicly accessible astronomical data archives that collectively make up what is now known as the Virtual Observatory (VO).

For all practical purposes, however, none of the limitations implied by the manufactured character of the images seems to constrain how that pool of knowledge resources is created or governed. Despite the uniqueness of each image, it remains part of a pool of knowledge that is neither exhaustible, rival, nor depletable. The nondepletable character of the relevant resource (the pool of image data) is central to the success of the Galaxy Zoo project. First, the night sky itself, of course, is an effectively inexhaustible resource. Second, the continuing SDSS venture and its associated data releases demonstrate that ever-improving observational, image processing, and data storage technologies are increasing the size of the knowledge pool more rapidly than astronomers can develop tools to classify and analyze it. Third, as the Galaxy Zoo project leaders learned, any particular image of a galaxy can be used or "consumed" many times over, and in fact the unexpectedly large number of classifications of the same galaxy by Galaxy Zoo volunteers turned out to be beneficial to the project as a whole. The increased number of classifications increased the level of confidence in the accuracy of the data as a whole (Lintott et al. 2008).

The sociotechnical detail surrounding SDSS is also relevant to the next cluster of questions, which goes to the default baseline or expectations regarding the shareability of that information, as well as to questions regarding how those resources might have been governed otherwise. Before turning to those questions in the next section, I note (with less detail) two other resource sets: Galaxy Zoo volunteers themselves, and the collected Galaxy Zoo data and associated scholarly papers.

The time and expertise of those who contributed to Galaxy Zoo is conventionally viewed as labor resources rather than as knowledge or information resources and is, therefore, correspondingly challenging to describe in terms of the commons framework. Yet it is precisely this pool that inspires interest in Galaxy Zoo as commons governance. Chris Lintott and Kevin Schawinski and other professional astronomers associated with creating and managing Galaxy Zoo and producing scholarship based on Galaxy Zoo data (that is, their scholarly colleagues, postdocs, and graduate students) were contributing and sharing their time and professional expertise with one another and with the broader astronomical research community much as any research scientists ordinarily would do. It is straightforward to draw a clear distinction between them and the corps of Galaxy Zoo volunteers, then to find a formal link between the expert contributions and those of the amateurs, given the fact that the latter were invited initially to conduct relatively elementary classification tasks and nothing more.

There is a meaningful sense, however, in which the time and expertise of Galaxy Zoo volunteers and experts alike may be viewed as information or knowledge resources as well as labor, so that governance questions become interesting at this second level. Consider, first, the time and expertise of Galaxy Zoo leaders. Most observers would reasonably conclude that the leaders' participation in the project involved sharing knowledge resources with each other and (potentially) with Galaxy Zoo volunteers. The leaders conceived of the project drawing on their astronomical expertise, designed it, and shared its concept, its execution, and aspects of the underlying expertise (some of which was explicit in the design of Galaxy Zoo, and some of which presumably was implicit) with the world of potential volunteers. Consider, next, the volunteers themselves. Galaxy Zoo conveyed knowledge sufficient for volunteers to effectively participate and invited participants to exercise human judgment. Having been invited to share their judgment as to relatively simple tasks, they were motivated to share their judgment—a kind of knowledge-based expertise. Through the Galaxy Zoo forum, they were given opportunities to share that judgment as to other, less-structured opportunities, such identifying and questioning the significance of celestial objects that volunteers could see alongside the galaxies. That expertise might not be characterized in terms of professional training and education, but the documented contributions of volunteers to the discovery of celestial objects beyond the galaxy classifications suggests that they were doing more than merely sharing clicks of a computer mouse. Some of those contributions led to published papers, suggesting that the professional astronomers recognized these volunteer contributions as more than mere thoughtless labor (Cardamone et al. 2009; Lintott et al. 2010).

Should thought*ful* labor be classified as a kind of knowledge resource? In terms of modern intellectual property orthodoxy, which is highly skeptical of any labor-based claims to proprietary rights or interests,[18] the answer might well be "no." But there is a significant difference between resources and legal rights governing those resources, and the former are the topic here. Omitting these resources from an account of the enterprise and treating Galaxy Zoo time and expertise simply in terms of a division of labor between professional and amateur astronomers likely misses some important aspects of what happened. There is abundant evidence of meaningful scholarly collaboration between and among the experts and the volunteers. The record also suggests that the professional astronomers went to some lengths to prevent disciplinary hierarchies from limiting forms of interpersonal interaction. With respect to human contributions, this was a knowledge-sharing enterprise as well as an exercise in distributing and sharing labor. The fit between the resource-character question and the nature of the shared contributions suggests, perhaps, that "knowledge resources" is an insufficiently broad question to be asking. In some respects, what was shared was not only a kind of astronomical judgment but also parts of what it means to be an astronomer or a scientist.[19] Akerlof & Kranton (2010) suggest that group and organizational identity may form an important part of an individual's contribution to organizational productivity but is missing from most economic accounts of organizations. In the case of Galaxy Zoo, identity, or identities, may have been shared, and they may have been among the most important "things" shared in terms of advancing the scientific goals of the project.

---

[18] See Feist Pub'ns, Inc. v. Rural Tel. Serv. Co., 499 U.S. 340 (1991) (rejecting labor-based "sweat of the brow" justifications for copyright interests).

Partially confirming that hypothesis is the project's pattern of scholarly publication and shared authorship. A number of Galaxy Zoo volunteers are listed as authors on scholarly publications to which they contributed. In the early papers produced by Galaxy Zoo leaders, the professional scientists tried to get journal editors to include as authors the names of all of the Galaxy Zoo volunteers (Schawinski interview 2011). When that approach was rejected, the leaders turned to acknowledging the volunteers in the paper and linking to an online image that includes all of their names (Burton 2012). The image can be viewed at http://zoo1.galaxyzoo.org/Volunteers.aspx. This sort of indirect credit might be viewed only as an effort to compensate volunteers in a way that motivates them to continue to contribute, but it is equally the case that the Galaxy Zoo leadership viewed (and views) the volunteers as genuine collaborators. Biagioli (2006) points out that scientific authorship should always be viewed as codifying contingent scientific norms. Authorship in one sense represents credit for having done some or all of the underlying work, but in a distinct sense it may represent accountability for that work, including errors, flaws, and possible harms that it causes. Galaxy Zoo's pattern of shared authorship neither represents full sharing of credit nor of responsibility, because of the physical and intellectual distance between the vast majority of the volunteers and the expert researchers who prepared the papers themselves. The expert author of the so-called "Green Peas" paper, disclosing and describing a class of objects identified initially by Galaxy Zoo volunteers, recounted the process by which the amateur observations were refined into a scholarly work product (Cardamone 2009; [Sheppard] 2009). At most shared authorship here represents a statement of descriptive and aspirational affinity between and among those who participated in Galaxy Zoo. When I interviewed Kevin Schawinski in his office (then at Yale University) and asked him to identify the members of the Galaxy Zoo team, he pointed to a poster of the Galaxy Zoo volunteer corps mounted on his office door, in effect saying "this is my team." The phrase that he did use, referring to the volunteers, was "direct collaborators" (Schawinski interview 2011).

Last, the Galaxy Zoo dataset itself should be considered as a kind of shared information resource and as a collection of shared resources. The time and expertise of Galaxy Zoo volunteers is not a well-defined shareable "thing" like an image of a galaxy or a galaxy itself or a pool of galaxy images, but via application of software and the volunteer's responses to classification questions (sometimes referred to as "votes") that volunteer's time and expertise is translated into a small smidgeon of metadata about that image. Each small piece of metadata is a tiny information "work" that is shared via the Galaxy Zoo database itself with all other, similar information works. Each such work in itself is of minimal value, but in the aggregate, millions of those works taken as a whole offer fertile ground for further research. The Galaxy Zoo database has been posted online, with all of the detail regarding individually contributed "works" intact, for full and open reuse by other scholars (Lintott et al. 2010).

---

[19] During 2009 and 2010, the Galaxy Zoo blog featured a series of posts framed as interviews with seven Galaxy Zoo experts and eight Galaxy Zoo volunteers, all women, titled "She's an Astronomer." (Masters 2011).

2. The Default Baseline: What Formal and Informal Expectations Attach to the Relevant Governable Resources?

This cluster of questions looks to distinguish between resources governed in the first place by default intellectual property regimes (patent law, copyright law, trade secret law, and the like) or by an absence of formal legal regimes (material that is in the public domain). The premise of both perspectives is that knowledge resources are nondepletable, and their nondepletable character is an important driver of commons creation and success. In the case of Galaxy Zoo, and continuing the theme introduced in the last section, the following suggests that both the nondepletable character of the relevant resources and the formal legal baseline for their governance may be less significant in practice than relevant community expectations regarding their creation and use.

The SDSS image data that constitutes the initial resources for Galaxy Zoo are potentially copyrightable. Although the images are meant to be absolutely faithful high resolution photographs of natural physical phenomena, it is plausible to suspect that they are characterized by the very modest amount of creative "authorship" that would place them within the scope of copyright. A host of individual researchers contributed to the process that produced the data; any or perhaps even all of them could claim standing, as a formal matter, as copyright "authors." Although the images were obtained by a collaborative that included funding and expertise from U.S. government agencies,[20] there is no suggestion that the images are U.S. government "works" that cannot be protected by copyright.

Functionally, the image data appears to have been treated by the SDSS and users of SDSS data as belonging to the kind of normative public domain of public information that we typically associate with scientific data. The Astrophysical Research Consortium published SDSS data to the world with the intent that it be used for follow-on research. The SDSS project website includes a restrictive notice to accompany any use of SDSS data: "Data from the SDSS public archive may not be used for any commercial publication or other commercial purpose except with explicit approval by the Astrophysical Research Consortium (ARC). Requests for such use should be directed to the ARC

---

[20] The Apache Point Observatory is operated by a consortium of research universities, listed at the observatory's home page at http://arc.apo.nmsu.edu/.

Corporate Office via ARC's Business Manager...."[21] Note a rhetorical ambiguity embedded in this statement that reinforces its public, shared character. The SDSS data archive is *public*, but it is presumptively *noncommercial*. One commentator has gone so far as to characterize the SDSS direct data release as a novel form of scholarly communication, because disclosing all of the data to the public contravened a scholarly convention that research data remain private or proprietary even so long as scholarly papers are published (Hey 2013). That treatment is consistent with the restrictive legend on the website. In effect, the SDSS data were simultaneously distributed and claimed as a shared scholarly resource.

With respect to Galaxy Zoo itself, the modest ambiguity did not cause any difficulty. The founders were aware of the restriction on use of the data from the beginning (Schawinski interview 2011). There is no evidence that the restriction was communicated formally to the broader Galaxy Zoo community; it seems to have been understood that the Galaxy Zoo project was by its nature a research program and therefore not commercial. There was never any question that SDSS image data could be freely shared within the Galaxy Zoo project and among Galaxy Zoo professionals and volunteers, even as that latter group expanded informally and significantly.

The other two pools of resources related to Galaxy Zoo, the knowledge pool that consists of the experts and volunteers themselves, and the Galaxy Zoo classification data, are somewhat easier to characterize. The leadership group, the professional astronomers, acted at virtually all times in ways that were consistent with "Mertonian" norms of scientific research: communalism (common or shared "ownership" of scientific discoveries); universalism (truth claims being evaluated according to objective or universal criteria); disinterestedness (selflessness); and organized skepticism (truth claims are subjected to community scrutiny) (Merton 1973). The two initiators, Lintott and Schwanski, engaged in what can only be characterized as a significant amount of entrepreneurship and promotion as part of getting the venture off the ground and sustaining interest in it (Lintott interview 2013). Press releases, interviews, and/or media appearances by one or both men were frequent (and remain so, in Lintott's case, with respect to the Zooniverse). Those efforts appear to have been motivated both by the desire to generate sufficient public interest in the project that it could succeed and possibly generate sufficient institutional credibility that it could retain funding from relevant supporters, particularly Johns Hopkins (host of the SDSS image archive) and Oxford (home of the servers with the Galaxy Zoo interface), and their respective public-funding sources. Interestingly, those promotional efforts, while important to the project's success, neither suggested any sort of proprietary claim to any Galaxy Zoo data nor compromised the outward association of the project with Mertonian norms. Hagstrom (1965) translated the Mertonian framework into an economic model framed as gift exchange. Scientific research was given to the broader community in exchange for nonmonetary recognition. It is fair to characterize the Galaxy Zoo team as operating in gift exchange mode, at least to a sizable extent.

---

[21] The notice appears at http://www.sdss.org/collaboration/credits.html, which has been unchanged since 2008. This page does not refer to copyright. There is no requirement that those who use or download SDSS data acknowledge the policy explicitly before they do so.

Likewise, the volunteer population appears to have expressed little interest in any non-Mertonian perspective on their efforts. They clearly welcomed the fact that their contributions were recognized in Galaxy Zoo's shared authorship practice. Individuals whose names were omitted from the list of the whole group pointed out the omissions, and they were corrected. Not recognized or pursued was the possibility that the volunteers could be regarded, in a class-based sense, as unpaid laborers doing the tedious work needed to enable the higher-profile, better-recognized scholarship of professional experts.[22] The modern phrase "citizen science" in this context replaces the nineteenth-century phrase "subordinate labourers," used by William Whewell (Cooper 2012b). With respect to the data itself, however, the leadership group understood that they had invited the public at large to participate in a science project, and the public understood that they accepted that invitation and thus subscribed to the implicit norms of science (Raddick et al. 2010).

Those two premises, as to the SDSS data and as to the efforts of the Galaxy Zoo team, merge in the character of the Galaxy Zoo classification database, which has been made available for public access. The default baseline with respect to that database is copyright law, and the database as a whole is almost certainly a copyrightable work of authorship. The Galaxy Zoo leadership developed software that aggregated the "votes" represented by individual classifications with respect to each galaxy, identified and segregated outliers, and translated the results into classifications in fact associated with each galaxy (Lintott et al. 2010). The result is undoubtedly and at least minimally creative. Galaxy Zoo pays forward the "no commercial use" restriction that accompanied the SDSS image data by associating the Galaxy Zoo data with a Creative Commons Attribution-Noncommercial-No Derivative Works 2.0 UK: England & Wales License. The Galaxy Zoo summary of its terms is worth quoting in full:

> The design and graphical elements of this site, including the Galaxy Zoo logo, are separately copyrighted and are not part of the Creative Commons license. Design and graphical elements © Galaxy Zoo 2007, All Rights Reserved.
>
> You may print, or download to a local hard disk, any content from this site for personal use. You may use any content from this site for personal or noncommercial use, with the exception of the Galaxy Zoo site design and logo, provided you acknowledge this site as the source of the content with the usage notice above. You may copy content to individual third parties for their personal or noncommercial use, but only if you distribute the content without modification, and acknowledge this site as the source with the usage notice above.

---

[22] The gendered dimension to Galaxy Zoo, as part of professional astronomy as a whole, did not go unnoticed (Masters 2011).

You may not, except with our express written permission, distribute or commercially exploit the content.

All the galaxy images that appear on the Galaxy Zoo site come from the Sloan Digital Sky Survey and are governed by its copyright policy. [Galaxy Zoo links here to the SDSS policy noted above, although what Galaxy Zoo characterizes as a copyright policy does not refer to copyright.]. Essentially, non-commercial personal, educational, or scientific use is granted as long as credit is given to the Sloan Digital Sky Survey, with a link to www.sdss.org. Specific acknowledgement must be given for non-commercial publication. To use the images for any commercial purposes, request permission according to the instructions in the SDSS copyright policy.[23]

Pause here to consider some technical copyright details. If, as this policy suggests, the Galaxy Zoo data comprise a single copyrighted work, then who is the author? One might look to the shared authorship practices of the group, which take a step in the direction of recognizing all Galaxy Zoo volunteers as contributors, if not formal, named authors. Acknowledgment does not an author make, but it should be recognized that between the shared authorship practice, the plausible treatment of each individual classification as a small work of authorship, and the reference by the Galaxy Zoo copyright policy to the SDSS restrictive notice, there is a gap. Individual volunteers are not asked or expected to license, waive, or assign any possible intellectual property interests that they might have in their contributions. SDSS image data are part of a formal IP rights scheme, and the Galaxy Zoo data are part of a formal IP rights scheme. But the volunteer contributions that are merged with the SDSS image data are not part of a formal IP rights scheme. In practice, this seems to matter not at all, and I do not mean to imply that there is anything defective about Galaxy Zoo's strategy. For reasons discussed below, Galaxy Zoo's approach here may be precisely the right one. One possible implication is that the formal IP baseline with respect to commons resources is less important than community or group expectations and how those expectations intersect with proprietary norms.

The other final output of Galaxy Zoo, scholarly papers, are subject to all of the usual expectations associated with scholarship in the astronomy and astrophysics domains. Virtually all of the papers that have used Galaxy Zoo data are, formally speaking, authored by leadership team members. As noted earlier, some papers have had nonteam members as named co-authors, and many have identified the population of Galaxy Zoo volunteers as contributors. But in Hagstrom's terms they, like the data itself, are a kind of gift. The papers are distributed publicly. In this context, e-prints are submitted to and posted on the arXiv website (http://www.archiv.org) that hosts open-access copies of papers in physics, including astrophysics, and some related disciplines.

---

[23] Copyright Notice, http://zoo1.galaxyzoo.org/Copyright.aspx, accessed June 12, 2013.

3. Goals and Objectives: What Commons Problems Are Being Solved?

The point of declaring that the governable resources are subject initially to an intellectual property regime, or are not, is to explore the reasons for the emergence of commons governance. What type or kind of social dilemma prompted a commons solution? The prototypical dilemma in a knowledge production and management context is framed by the formal "tragic commons" hypothesis, that in the absence of individuated, marketable proprietary rights, a resource would be overconsumed. An alternative dilemma posits that the presence of proprietary rights themselves with respect to a given resource might give rise to a kind of coordination or transactions costs problem, if multiple rights and/or owners need to be combined in order for the resource to be exploited effectively. The prototypical response to such a dilemma is to preserve the resource as a formally and fully open or public thing. Commons governance is suggested as a method by which either of these dilemmas might be solved by enterprise participants exiting the default market-based system of proprietary rights, or the default open and public system, and working instead through a blend of formal and informal or norm-based rights and interests.

In the case of Galaxy Zoo, it might be said that neither sort of dilemma directed the commons solution, although the foundation of the galaxy classification problem was, in a sense, an underproduction problem framed in terms of coordination of participants at scale. How could all of those galaxies be studied and a suitable database of metadata created? Classic law-and-economics theory might suggest creating a market defined by transactions in individual interests in particular "votes" or classifications. One need not imagine some kind of public auction or trading floor; the "votes" could be seamlessly and automatically packaged and transferred, with an accompanying transfer of rights, in an online space little different from Galaxy Zoo itself. In practice, as noted above, that theoretical framework did not guide the development of Galaxy Zoo. Lintott and Schawinski did not start from scratch; they started from a position of commons to begin with (SDSS scientific data as a shared communal resource), and they adapted from that point forward. That is, if the discipline of professional astronomy constitutes commons governance with respect to scientific research in the first place, as it likely does, then the challenge that Lintott and Schawinski faced when they needed to classify their SDSS image data was a matter of choosing (1) to maintain the existing professional commons model and coding the data themselves; (2) to abandon the commons model in whole or in part, either by obtaining resources from a public authority, such as coding services or money to hire coders, or by organizing a market of coders with proprietary claims to exchange and housing the market entirely within their enterprise (one could characterize this approach either as a professional commons model accompanied by a clear division of labor, or as abandonment of the commons model and adoption of a firm-based model); or (3) to expand commons governance to include more "commoners."

In effect, this last option, to expand commons governance, is the strategy chosen by Galaxy Zoo. Commons was built adjacent to commons. The line between "building a firm of coders" based on division of labor, on the one hand, and "including more commoners," on the other hand, appears to be a thin one in theory. One person's commoners are another's (human) computers (Fortson et al. 2009). In the first half of the twentieth century, the term "computer" referred to individuals, usually women, who were hired to classify large batches of observational astronomical data (Nelson 2008). The affinity between those computers and contemporary Galaxy Zoo volunteers cannot be overlooked. But in practice, that affinity seems to have mattered less than the distinction. The volunteers believed that they were part of the scientific community, rather than subservient to it.

4. Resource and Community Openness: How Open Are the Relevant Resources and Groups?

"Openness" is not a simple binary, but openness of some sort is part of commons governance. Franzoni and Sauermann, in their recent discussion of "crowdsourced" scientific research, including Galaxy Zoo, focus particular attention on dimensions of openness (2014: 7-10). The idea that knowledge or information resources are shared depends to some extent on their not being fully controlled or controllable by a single agent or individual. The related groups of resources and communities described above operate with different degrees of openness, both as to membership and as to the creation and use of resources.

The pool of SDSS image data can only be added to by the scientific professionals charged with administering the instrument in New Mexico and its associated hardware, software, and database, but in theory it can be used by anyone, once the data is released publicly. In practice, of course, the data set is so large and so complex that few people will be able to make meaningful use of it. That very problem gave rise to Galaxy Zoo itself. The Galaxy Zoo community includes at least three obvious levels of inclusiveness. With some nuance added, several more are revealed. The leadership team of expert astronomers is relatively closed; over time, some leaders have joined and some have departed, but typical scientific and institutional credentials are required to become an active member of that group. The population of Zooite volunteers is open to anyone with access to an Internet connection; in practice anyone *can* (or could) classify galaxies. In between is the population of volunteers who are more active participants in forum discussions both about galaxies and about other astronomical phenomena revealed in the SDSS image data. Registration is all that is required (pseudonymous registration is acceptable); no one is required to meet any quantity or quality threshold in order to remain a Galaxy Zoo or forum member in good standing.

In addition to those three levels of community openness, more subtle distinctions appear among Galaxy Zoo volunteers. Some actively collaborate with professional astronomers in authoring scholarly papers; access to that status is relatively open and depends on substantive contributions to relevant forum threads. Others have taken leading roles in structuring and monitoring message boards on the forum. That status appears to derive from informal arrangements and relationships with the Galaxy Zoo team, particularly with Chris Lintott (Lintott interview 2013) rather than from any formal selection or election process. It corresponds roughly to length and depth of engagement with the forum. As with many Internet forums, more active and more thoughtful participation leads to more substantial responsibilities. While the pool of SDSS data themselves is relatively fixed, with new data added only by professionals and existing data only analyzed by volunteers, the pool of volunteers is quite fluid, especially in the middle and bottom (largest) populations. Leadership and coordination is limited largely but not exclusively to the professional astronomers.

Last, there is the pool of Galaxy Zoo classification data, which has been published (via http://data.galaxyzoo.org/) and which remains available to follow-on researchers. A relevant nuance here confirms the claim above that "openness" is not a simple binary. The Galaxy Zoo leadership team held off releasing the underlying data until after publication of the initial Galaxy Zoo scholarly papers (Lintott interview 2013). The delay neither added to nor detracted from the mechanics of Galaxy Zoo as a scientific project. Instead, it followed from the need to coordinate the novel governance dimensions of Galaxy Zoo with existing imperatives of scholarship and professional advancement by the leadership group.

It is not clear from the description above that "openness" as a separate inquiry adds much to the analysis of Galaxy Zoo, once the sociotechnical dynamics of the enterprise are accounted for. "Open" commons governance does not add resources at the beginning of the process and accounts only for a part of the collaborative work that goes into producing scientific papers. There is no extraction and replenishment process with respect to the knowledge resources involved aside from expansion and management of the pool of Galaxy Zoo volunteers themselves. Membership in the Galaxy Zoo project exists at various levels; the complete openness of the classification project to volunteer contributions means that there is little meaningful boundary between "inside" and "outside" the project. The key variable, already discussed, is the manner in which contributions are coordinated.

5. "Rules-in-Use" (Narratives; Appropriation, Management, and Replenishment; Institutional Nesting; Relevant Legal Regimes; Discipline)

In terms analogous to those set by Ostrom with respect to the IAD framework, Galaxy Zoo is both a resource institution and a kind of "action arena" in which group members interact with resources according to informal and formal rules. Once the system is set up, in other words, then it operates. In practice, of course, distinguishing the elements of the mechanism from its operation is more art than science. This section, however, attempts some of that task.

- *Narrative(s)*
  The stories that Galaxy Zoo tells about itself, and that are told about Galaxy Zoo, do at least two things. In part they frame the analysis here, and in part they frame how the enterprise understands itself. The discussion in the introduction to this chapter disclosed much of the relevant narrative of Galaxy Zoo. Commentary both internal to the Galaxy Zoo project (such as articles written and interviews given by Galaxy Zoo members) and external to it (articles written by outsiders, either for scientific publications or for the popular media) appear to emphasize "citizen science," "crowdsourcing," and "Big Data" or "eScience" if not in equal measure then at least to a significant extent in each case. The common thread is "science," and within science, "astronomy." Galaxy Zoo is simultaneously characterized as an extraordinary solution to an extraordinary problem and as part and parcel of scientific traditions. That duality seems to be critical in keeping all aspects of Galaxy Zoo functioning well.

As with many extraordinary things, in other words, evidence of continuity is as compelling as evidence of innovation and change. The underlying scientific problem, the classification of galaxies in order to understand their dynamics, is a long-standing challenge for astronomers and was among the questions addressed by Edwin Hubble, for whom the space telescope is named. Amateur support for astronomy, if not necessarily "citizen science," is among the oldest traditions in research science. Galileo, Newton, and Kepler were amateurs in their own times. Patterns of broad amateur participation in observational astronomy have been well documented for decades (Berensden 2005; Ferris 2002; Lankford 1981). In many respects Galaxy Zoo is a broadening and deepening of explicit, long-standing amateur interests, including those of the leaders themselves (Lintott interview 2013). When Lintott and Schawinski invited the public to help them classify galaxies, it is likely that those who responded included many people with existing, if often latent, interests in the field (Raddick et al. 2010). Galaxy Zoo likewise offers a robust platform for teaching lay and youth audiences about astronomy.[24]

- *Appropriation and replenishment processes and rules*
  As to the SDSS image data and related sky survey data, the sky and the associated imaging data are functionally nondepletable. We do not yet know the limits of the universe, and advancements in sensing and detection technologies mean that the quality and quantity of image data continues to improve. Replenishment of the classification data, either by permitting multiple classifications of the same galaxy or by putting forward additional sets of data images for classification (as in Galaxy Zoo: 2 and Galaxy Zoo: Hubble) has proved to be robust in practice, despite the absence of any formal or informal rules that govern how the classification data is created, such as how many times a given galaxy may be classified, or how many galaxies may be classified by a given volunteer.

---

[24] The Zooniverse includes its own set of resources for teachers. See https://www.zooniverse.org/education and http://www.zooteach.org/zoo/galaxy_zoo, both accessed Sept. 10, 2013.

"Depletion" and "replenishment" and accompanying limits and duties are not quite accurate descriptions of the operation of the data analysis project that defined Galaxy Zoo in the first place. The SDSS image data simply "is," although its parameters expanded and continue to expand. The classification data accumulated through repeated "votes" by Galaxy Zoo volunteers, although no additional contributions were expected from existing members. On the one hand, older data did not decay or disappear, so there was no "depletion." On the other hand, new data did not simply add to the existing archive on a one-vote-per-entry basis; the relevant resource was supplemented, or enhanced, but not "replenished." As Lintott explained when the Galaxy Zoo data were released, a complex computer algorithm was developed to convert raw votes into classification data (Lintott et al. 2010).

Moreover, the Galaxy Zoo community identified and pursued additional celestial objects for study, which means that a certain type of resource creation and/or replenishment was possible even while the original Galaxy Zoo task entailed classifying a fixed data set. A technical platform built for one purpose turned out to support a community that was capable of and interested in pursuing many more tasks. Galaxy Zoo facilitated research into additional objects via a series of technical and social decisions. First, the Galaxy Zoo website classification interface included a link next to each image to the Sky Server Object Explorer for that image. Volunteers were not limited narrowly to the classification exercise with respect to each galaxy but had the power, if they had the imagination, to explore further. Second, the creation and support of the Galaxy Zoo forum offered those more imaginative and curious volunteers a venue for sharing their questions and insights, so that questions and possible research about these additional objects could ferment. Third, the Galaxy Zoo team quickly accepted the productive contributions of volunteers and collaborated with the volunteers, rather than directing volunteer time and energy solely to the classification task. Galaxy Zoo was productive commons governance in several senses, not the least of which was the creation of new knowledge beyond what the organizers planned for.

- *Nesting*

Is Galaxy Zoo or any of its constituent parts nested in higher order commons governance regimes? The relationships among the different shared sets of data in Galaxy Zoo initially appear to be mostly linear. SDSS image data were produced. Then, in response to particular research problems, a broad pool of human knowledge resources were brought to bear on data analysis, leading to the creation of a database of metadata about the original data set, to scholarly publications interpreting that metadata, and to new observations and scholarly publications made possible in large part by the construction of the pool of human resources.

That linear characterization is fair in one sense, but the self-reported descriptions of Galaxy Zoo as a collaborative, iterative enterprise should also be taken seriously. Galaxy Zoo both produced new astronomical knowledge and also produced itself. In a broad sense, the creation and disciplinary acceptance of Galaxy Zoo as a research enterprise (via the addition of new team members and via publication of its papers in peer-reviewed journals) seems to be linked to its origins both within the research university setting (Oxford, to begin with, with immediate and direct links to several other universities, including Johns Hopkins and Yale) and within the communities of scientific research generally and of astronomical research in particular. It is noteworthy that following his Oxford experience and after he co-initiated Galaxy Zoo, Schawinski was awarded a postdoc at Yale, which houses a highly regarded astronomy and astrophysics program, and that Yale warmly welcomed Galaxy Zoo as part of his research agenda (Schawinski interview 2011).[25] Each of those things may be fairly characterized as commons governance in their own rights and therefore as material contributors to the success of the Galaxy Zoo as commons itself.

- *Relevant legal regimes and discipline*

I have not found any legal regimes that bear directly on any features of Galaxy Zoo. The forum is nominally subject to the "safe harbors" for hosted content described in Section 512 of the American Digital Millennium Copyright Act.[26] The image data is now hosted on a commercial third-party cloud storage service, which is subject to both Section 512 of the DMCA and to Section 230 of the Communications Decency Act, which immunizes hosts of that sort from liability for non-IP claims associated with merely hosting the data. But there have been no reported episodes or incidents of bad behavior taking place on the forum that rise to the level of implicating possible legal claims, and there are no reports of moderators having had to intervene on disciplinary grounds other than very infrequently (Schawinski interview 2011). The forum moderators are relatively strict with respect to the use of inappropriate or vulgar language on the forum, in order to ensure that the forum is welcoming and friendly to all comers, including children (Lintott interview 2013).

The disciplinary question is independently noteworthy. There appears to be no formal or informal disciplinary mechanisms with respect to any of the commons communities included within Galaxy Zoo. Nor is there any formal organizational hierarchy within the leadership team (Lintott interview 2013; Schawinski interview 2011). Given the size and presumed diversity of the volunteer population, that seems remarkable, though perhaps it is less so when one considers the relatively narrow scope of "community" participation, at least with respect to classification. The scope of bad behavior available to classifiers is pretty small; individuals could intentionally mis-describe an object. But the algorithm that translates such "votes" into classifications is set up to anticipate and discard outliers, and there is little reason *ex ante* to prescribe a regime for excluding those individuals from continued participation. In other words, Lintott set himself the initial challenge in 2007 of figuring out how to find *enough* volunteers (Lintott interview 2013), rather than to find the *right* volunteers.

---

[25] Perhaps coincidentally, the chair of the Yale Astronomy Department at the time was Meg Urry, who is one of a relatively small number of senior women in professional astronomy (Masters 2010).

[26] 17 U.S.C. § 512 (2012).

Discipline, to the extent that it exists, is either technical, as just described, or social. Volunteers who did no more than classify galaxies would never trip disciplinary wires set up to catch outlying votes; those who did register for the Galaxy Zoo forum and posted there would be subject to moderation. The fact that little moderation has been required suggests that despite the diversity of background that characterized the volunteers, in fact they mostly shared a commitment to the broad ambitions of the project. I hesitate to characterize that commitment in terms of social norms or norms of any kind, because that term sometimes suggests the existence of a kind of informal rule-set. In the Galaxy Zoo context, the project was characterized by a strong alignment of interests, expectations, and goals both at a high level ("science" and "astronomy") and at ground level (investigating galaxies and black holes).

6. Outcomes and Assessment

In terms of the number of volunteer participants, their passion and focus, the amount of morphological classification data and other scientific data generated, and, importantly, the number of scholarly papers published using that data, there seems to be little reason to question the success of the Galaxy Zoo project, both on its initial terms and on the terms that evolved over time, with the emergence of the forum and with volunteer-initiated discoveries. While its absolute success seems clear, its relative success is less obvious. The next section contrasts Galaxy Zoo's governance, which I characterize as commons governance, with possible alternatives.

B. INSTITUTIONAL ALTERNATIVES

Fully appreciating the institutional character of Galaxy Zoo as commons requires a brief comparison between Galaxy Zoo's governance mechanisms and the most plausible alternatives. How is Galaxy Zoo different, and in what ways have those differences benefited or imposed costs on Galaxy Zoo?

From the standpoint of knowledge and information management, three possible alternatives seem plausible.

The first looks at the galaxy classification exercise as an exemplary scientific research project undertaken by a graduate student or by a faculty researcher. It would represent professional science, rather than citizen science. Under this model, a PI (principal investigator) would undertake responsibility for all of the relevant data acquisition and analysis, alone or by supervising a team of research assistants. Classification would take place by hand, given the judgment early on in the case of Galaxy Zoo that computer algorithms could not perform the task adequately. Given the Kevin-week unit of analysis developed using the manual classification of galaxies by Kevin Schawinski himself, it seems likely that this hypothetical manually created Galaxy Zoo dataset would be inferior to the actual Galaxy Zoo dataset in one or more ways. Fewer galaxies would be classified (and therefore fewer of the target galaxies might be identified), or the classification would be less accurate (because fewer galaxies would be classified multiple times), or the project would take much, much longer to complete and fewer research findings would be produced over the same time span. This alternative construction would lack the spillover benefits that accumulated as a result of exploring objects through the SDSS Object Viewer and the formation of the Galaxy Zoo forum.

A second alternative construction borrows from well-known examples of peer production often characterized as commons: open source software development and Wikipedia. Both Wikipedia and an open source software development project operate approximately as Galaxy Zoo does. A very large number of contributors who may be largely anonymous or pseudonymous contribute small bits of labor and knowledge to a product produced collectively and that is much larger than any single person, or even any smallish group of people, could produce on their own. In the case of Wikipedia, this is the Wikipedia online encyclopedia itself; in the case of an open source software development project, it is the resulting computer program. In both cases the community of participants is itself a kind of ongoing product. In the context of open source projects, Kelty refers to this phenomenon as a "recursive public" (Kelty 2008). What he means is that the social collective both constitutes and is constituted by the product that it produces via knowledge and information sharing. Wikipedia, though not the subject of Kelty's study, bears many of the same hallmarks. The community of Galaxy Zoo Zooites and zookeepers and the database and scholarship that they jointly produce appears to have much in common with the "recursive public" model. Galaxy Zoo is both process and product.

There is at least one key distinction, and that has to do with some details of the governance strategy used by both Wikipedia and open source projects. Open source projects and products are defined in part by their formal software licenses. Each license is an essential part of the open source knowledge governance strategy, ensuring that the knowledge pooled within each project is shared both with active developers and with users and casual modifiers of the program (Madison 2003). Licenses vary in their details, but open source licenses all require that the source code for the program be made available to users and future developers so that they might modify it. In many cases the license requires that users and future developers agree that they will likewise share the source code to their modifications of the program. This duty to share is often a strong informal, normative commitment of open source communities, but it is also formally documented. Each member of an open source community, which is to say, each user of a particular open source computer program, agrees to certain formal terms and conditions that accompany access to that program, and among those terms, typically, is the duty to license back to other users (to share) modifications made by that user. From time to time, when disputes arise about the obligations of users of an open source program, the formal document is the basis for resolving the dispute,[27] and as matter of best practices, it is possible to trace individual contributions and their compliance with licensing obligations. Open source software projects are thus paradigmatic examples of commons governance, but the openness and shared character of the open source project is both norm-based and legally and technologically formal.

Likewise, contributors to Wikipedia often operate under a strong normative commitment to openness, but they, too, must formally subscribe to a form of license (a Creative Commons license) that attaches to each of their contributions, no matter how small, and ensures that those contributions will be available to future Wikipedia users on open terms. When individual contributions are aggregated within particular Wikipedia entries, the underlying Creative Commons licensing that attaches to those contributions is, for practical purposes, pooled as well, so that entire Wikipedia entries are made available to users under Creative Commons licenses. The underlying individual licenses and entries, while fully available in detail to those who would look under Wikipedia's hood to determine how the end product was created, are hidden from immediate view. Wikipedia is thus likewise a paradigmatic example of commons governance (Benkler 2006: 70–74), but like open source software projects, its openness and commitment to knowledge sharing is defined formally, in both technical and legal terms, as well as normatively.

By contrast, Galaxy Zoo (and for that matter, all of the Zooniverse projects) required no equivalent formal license or assignment of legal rights in connection with individual contributions and classifications, and those contributions are far less individuated and traceable in the context of the overall Galaxy Zoo classification database. A number of questions arise. Why does Galaxy Zoo lack the formality of an open source project or Wikipedia? Could Galaxy Zoo have been organized as either of these were? Has Galaxy Zoo's lack of that formality interfered with its development in any way? Could it be said that Galaxy Zoo has succeeded in part because of the lack of formality? If so, then why, and how?

Definitive answers are not possible, but a number of considerations seem to be at work. In the first place, contributions to an open source software project and to Wikipedia, even if trivially small, are arguably protected by applicable copyright law. At the least, it is not implausible under current law that should a dispute arise among Wikipedians or among open source software developers, a court would recognize a copyright interest in an individual contributor/claimant. If that is the case, then developing and using a license to govern knowledge contribution and use seems wise, out of an abundance of caution. The standardized license minimizes a specific type of transactions cost that would otherwise be associated with clearing all rights from individual participants and that would be associated, in particular, with a disruptive outlier who refuses to play by informal norms. By contrast, it is less clear (though as I noted above, in my view not implausible) that galaxy classifications in the Galaxy Zoo project could be treated as copyrightable works of authorship for copyright purposes rather than as contributions of time and effort. Legally speaking, even a cautious copyright lawyer might see no reason for concern. In the second place, both open source software projects and Wikipedia likely have greater needs to for what we might call "lateral" coordination of the contributions of individual programmers and authors/editors. In technical terms, a modification of a computer program must work effectively with the existing program; an edit to a Wikipedia entry must relate in some coherent way to its existing content. The license does not ensure that either of these things will happen, but against a background norm of copyright status for each contribution and each entry, the license permits others—higher level technical coordinators and editors—to have the access to the work that they need in order to make it happen.

---

[27] See Jacobsen v. Katzer, 535 F.3d 1373 (Fed. Cir. 2008).

Galaxy Zoo is different in key respects. Each contributor's classifications simply go into the database of classifications. Coordination among the classifications is in effect "vertical," in the sense that further algorithmic work is done to analyze the classifications as a group (including, among other things, identifying and distinguishing outliers), but there is no equivalent coordinating "use" of each classification relative to existing classifications. Moreover, and perhaps more important, for reasons and in ways reviewed above, Galaxy Zoo seems to have successfully cultivated an ethos of "science" among its participants, so that the normative constraint on both data creation (everyone will do their best) and on dispute resolution (no one will claim individuated rights in a particular classification that could hold up the productive use of the pooled data) seems to be more than adequate. One cannot say for certain, but the disciplinary idea of "astronomy" seems to hold people together in Galaxy Zoo well enough, as a governance strategy.

Two further details should be noted, both of which suggest that the conclusion of the last paragraph may be on the right track. First, in the domain of commercializable "User Generated Content," or manufacturing platforms that invite end users to share ideas with for-profit companies that may be built into new versions of commercial products, it is typical for those platforms to offer mandatory terms of use by which each end user assigns all relevant rights to the manufacturer in the event that an individual contribution, which may or may not rise to the level of copyrightable authorship protected by applicable law, makes its way into a real product.[28]

Earlier in this chapter, I noted the absence of such a fail-safe provision and noted that it might be part of a wise strategy on the part of Galaxy Zoo. Here is why. Its absence may be traced in part to the fact that for all practical purposes, there are no commercial implications of individual contributions or of the Galaxy Zoo classification database as a whole. Yet Galaxy Zoo has attached a no-commercial-use Creative Commons license to its own database. So the absence or presence of a formal governance document in the Galaxy Zoo context is not entirely due to the presence or absence of plausible commercial potential.[29] More important, Galaxy Zoo contributors are permitted to submit more than simply the classifications that become part of the Galaxy Zoo database; over time they have contributed additional scientific observations and analysis. There is little doubt that the form of this material is potentially copyrightable, and that its substance may end up in copyrightable material (the research papers) and has ended up there, relatively directly. Yet registered users of the forum are not required to acknowledge any formal restriction, release, or license of material shared there. I conclude that the practice of inclusiveness in terms of shared authorship for Galaxy Zoo papers, coupled with the practice of inclusiveness with respect to participation in Galaxy Zoo generally, has been sufficient to support the project's relative informal governance. Formalizing rules regarding contributions and authorship with respect to forum participants might have had unanticipated harmful consequences, by undermining the normative benefits of associating the entire enterprise and its participants with astronomy. A formal license or release may have signaled that Zooites were not fully welcome as citizen scientists, or that their potentially valuable contributions were being appropriated for less than fair value in return (even in Hagstrom's sense of scientific research as gift exchange)—in other words, that this was not only research science. Either signal could have reduced participation levels or reduced the positive spillover observed in practice. To be sure, either result is speculative.

---

[28] See, e.g., the LEGO CUUSSOO User Innovation Platform Terms of Service, http://lego.cuusoo.com/terms, accessed June 13, 2013.

In sum, it seems clear that Galaxy Zoo could have added a layer of formality (a license or release or both) either to the classification exercise or to the forum, or to both, very much as Wikipedia and open source software projects have done. For the reasons surveyed above, doing so likely was not necessary. The lesson is that peer production comes in many forms. Here, peer-produced "astronomy" itself is a kind of commons.

The third plausible institutional alternative borrows not from a hypothetical research project but from an actual one: the Nearby Supernova Factory. If the first alternative is grounded in traditional scientific organization and the second is grounded in newer forms of peer production, then this third alternative seems to be grounded most plausibly in conventional understandings of the firm. I use this particular example because like Galaxy Zoo, it is an organizational form situated in data-intensive science, particularly astronomy and astrophysics.

The Nearby Supernova Factory (SNfactory), housed at http://snfactory.lbl.gov/, is a formal interinstitutional astrophysics data collection, curation, and distribution enterprise. It is an international astrophysics experiment "designed to collect data on more Type Ia supernovae than have ever been studied in a single project before, and in so doing, to answer some fundamental questions about the nature of the universe."[30] Specifically, it is "an experiment to develop Type Ia supernovae as tools to measure the expansion history of the Universe and explore the nature of Dark Energy. It is the largest data volume supernova search currently in operation."[31] Planned in 2001 and launched in 2002, SNfactory has six participating institutions (three in France, two in the United States, and one in Germany), and several dozen participating individual members, about half of whom are in the United States and the other half in France (a very small number of members are located in other countries). Membership is interdisciplinary; it includes physicists, scientists, and software engineers, among others.

---

[29] Trademark concerns seem to be much more real, however, and Galaxy Zoo's copyright notice takes care to (try to) forbid commercial use of Galaxy Zoo's logo and site design.

The project uses a CCD camera attached to its primary telescope, originally at the Palomar Observatory near San Diego, California, and now located in Chile, to collect up to 80 GB of data each night, using specifications provided by a geographically distributed group of two to six people.[32] That data becomes part of SUNFALL (SUperNova Assembly Line), "a collaborative visual analytics software system to provide distributed access, management, visualization, and analysis of supernova data."[33] The data is transferred via a high-speed network from Hawaii and Palomar to the Lawrence Berkeley National Laboratory (LBNL), where the data is stored in a giant database.[34] Follow-up spectroscopic screening and analysis takes place at LBNL, at facilities in France, and at Yale (Hules 2011). The search for supernovae, the task of analyzing them to determine whether they are Type Ia supernovae, and addressing the problem of understanding their brightness (a key measure in using Type Ia supernovae as cosmological tools, that is, to understand the rate of expansion of the universe), are undertaken by researchers elsewhere, particularly at Yale.

In sum, the "factory" production process involves three steps. First, broad scans of the sky are obtained. That is the function of the discovery camera, originally at Palomar. Second, specific coordinates of candidate objects are fed through the software to the camera; repeated observations of these objects are important in order to obtain data regarding changes in brightness. The images themselves are processed using spectrographic analysis, and eventually the resulting data is interpreted by human researchers. The project name—the "Factory"—owes its origin to the fact that the task is to produce new "things": so-called "standard candles," or cosmological objects whose brightness can be standardized and therefore used in cosmological analysis.

The distributed, interdisciplinary data curation and management strategies of SNfactory are integral to the project and key contributors to its success. SNfactory and SUNFALL have reduced false supernovae identification by 40 percent, improved scanning and vetting times by 70 percent, and reduced labor for search and scanning from six to eight people working four hours per day to one person working one hour per day. The project has led to ten publications in 2009 in both computer science and physics journals (Hey 2011). It has been able to determine the intrinsic standard brightness of Type Ia supernovae with much improved accuracy. The project has evolved somewhat since its origins (the relocation of the camera suggests as much), and its success has spawned some rivals, such as the Palomar Transient Factory, at Caltech, housed at http://ptf.caltech.edu/iptf/.

---

[30] About the SNfactory, http://snfactory.lbl.gov/snf/snf-about.html.
[31] The Nearby Supernova Factory, http://snfactory.lbl.gov/.
[32] The instrument was originally located at the W. M. Keck Observatory in Hawaii but was relocated to Palomar and has since been relocated to Cerro Tololo, in Chile. The coordinates used by the camera are now generated via computer software.
[33] SUNFALL: SUperNova Factory Assembly Line, http://snfactory.lbl.gov/snf/snf-sunfall.html.
[34] The storage facilities are provided by the National Energy Research Scientific Computing Center, or NERSC.

As a large-scale, distributed, information technology-based enterprise intended to manage a large volume of astrophysical data, the SNfactory has much in common with Galaxy Zoo.

Yet in important respects, the SNfactory governs itself and its data differently, and the differences are instructive. In effect, the SNfactory is a hierarchical firm, producing scientific products (its standard candles) and otherwise operating, with respect to its data, within the bounds of astrophysics as a professional discipline. Membership in the SNfactory is clearly and publicly defined and limited to the institutional homes of participating researchers. These are defined in the first instance by the several institutions that are listed sponsors of the project. Individual researchers who are members of the SNfactory project are employed by these institutions or are students of faculty and researchers employed there (or both) and are academic scientists from a variety of disciplines: astrophysicists, computer scientists, and engineers from several engineering subfields (including electrical, mechanical, and optical). The physicists work with the data, and the computer scientists and engineers are responsible primarily for design and maintenance of the hardware and software facilities used to analyze the image data. The latter are referred to by the project as "builders" rather than as "members."

In contrast to Galaxy Zoo's informal and undocumented collaborative of zookeepers/leaders and Zooites, the organizational structure of SNfactory is part of a formal, documented governance structure.[35] The SNfactory is managed by the SNfactory Collaboration Board (SCB), which consists of one representative of each sponsoring group, plus the Principal Investigator of the Supernova Cosmology Project at Lawrence Berkeley National Laboratory, a spokesperson for the participating French research consortium, and a project manager. Operation of the project is delegated by the Collaboration Board to an Operations Committee (OC) (a small group of member researchers, and the project manager). Individual graduate students and postdocs can be added to the project at the discretion of the leadership of each participating group. The Collaboration Board must approve admission of new faculty researchers or permanent staff.

---

[35] The formal governance document is posted publicly at http://snfactory.in2p3.fr/people/SNFactory_Organization_v4.1.pdf, accessed June 13, 2013.

Equivalent formality attaches to the SNfactory's approach to governing the shared data itself. As is common in scientific research, and as is common in for-profit firms that rely on confidential technical information or know-how (that is, trade secrets), the enterprise draws a clear and explicit line between sharing data among members of the collaborative, which is expressly permitted, and sharing it with outsiders, which is forbidden except with the permission of the Collaboration Board. More of the collaboration document is dedicated to identifying express standards related to publication of works based on SNfactory data. Where the Galaxy Zoo team informally adopted a shared authorship norm that attributed Galaxy Zoo scholarship to all Galaxy Zoo participants, the SNfactory reached an equivalent conclusion by formal rule: "Any paper written by a collaboration member that uses data, software, or internal group knowledge that comes out of the collaboration's work is assumed to be a collaboration paper unless otherwise agreed to in advance by the SNfactory Collaboration Board."

Whether or not all of this formality is followed strictly in practice, the only thing that surprises about this governance approach is the fact that it has been reduced to writing in such a formal way. The content of the rules is entirely consistent with underlying practice in scientific research and with a linear "production line" concept. There are well-defined roles and responsibilities among collaborating researchers and their institutions, and a firm injunction against disclosing unpublished data without the consent of the team. (That injunction likes serves the entirely ordinary scientific interest in avoiding premature publication of incompletely analyzed data, an outcome that could jeopardize institutions, careers, and funding; the Galaxy Zoo team informally adopted a similar approach.) The sociotechnical architecture that secures this model extends to the data itself. Whereas Galaxy Zoo used publicly accessible data and shared its data publicly rather than via a scholarly platform, SNfactory created its own data and managed access to its by network security protocols, that is, by passwords and an addressing protocol that renders the relevant website hidden from Google and other search engines.

To an observer outside the disciplines relevant to the SNfactory, perhaps the most significant thing about the operation of the project is its very normalcy. In effect, the narrative might explain, this is how science is produced. Within the astrophysics literature, the SNfactory is presented as a technical solution to a very important new scientific problem: understanding the properties of "dark energy," which is now believed to be the key to measuring the rate of expansion of the universe. The collaborative elements of the project and its governance structure are largely hidden from public view. They are not hidden outright, but they appear to be treated merely as scientific or research infrastructure (which, of course, they are). The narratives that accompany Galaxy Zoo stress its extraordinary character. Galaxy Zoo is treated as a novel form of scientific collaboration.

The comparative institutional question is whether it is conceivable that Galaxy Zoo might have been organized roughly as SNfactory has been organized, as a large, complex but ultimately ordinary form of formal hierarchical collaboration among scientific researchers. The scientific problems addressed by SNfactory and Galaxy Zoo are not identical, but they are closely related. In both cases, scientists asked how to coordinate analysis of a massive amount of data about the universe. It is conceivable that institutions to classify galaxies might have been organized in more conventional hierarchical terms, and Galaxy Zoo itself is hardly without its important hierarchies. But it seems doubtful that Galaxy Zoo could have achieved its levels of success by building in the kinds of organizational formality that characterize SNfactory and the sense of proprietary interest in the raw data that specifically concerns that project. Like many if not most scientific research collaboratives, SNfactory is characterized by an acute sensibility with respect to who is "in" the collaborative and who is "out." Galaxy Zoo, by contrast, never imposed such a bright-line boundary, and in many respects went out of its way to express a participatory norm that was both extremely inclusive and that projected the idea that all Galaxy Zoo volunteers were, in effect, part of the scientific team. Even if that idea took hold over time rather than immediately upon the launch of the project, the evidence suggests that it played an important role in encouraging volunteers to participate both in the classification exercise and in the Galaxy Zoo forum.

In sum, if a scientific discipline and scientific research are both forms of commons governance, then the demands of scale imposed by data-intensive science may be satisfied in multiple ways. As in SNfactory, with highly specialized data-processing needs, related commons may be joined in a kind of hierarchical data-processing firm, or, as in Galaxy Zoo, with much simpler data-processing needs. An existing commons may be expanded in scale so that, functionally, almost anyone who wishes to join may do so—so long as the individual plays by disciplinary rules. Of course, a full assessment of the costs and benefits of each approach should account not only for the community governance challenges in each setting and the relative complexity of their data analysis challenges but also for the respective spillovers associated with each. In the case of the SNfactory, spillover effects, if any exist, are likely focused on improving the expertise and impact of the associated researchers and their institutions in connection with this particular project. Progress in the field is achieved as it often is in science, via rival teams of researchers trying to outperform each other. In the case of Galaxy Zoo, spillover effects included unanticipated research findings achieved by or together with volunteer researchers, expansion of the field as volunteers entered academic astronomy, redirection of research trajectories as Galaxy Zoo team leaders expanded and redirected their technical platform to address new research challenges, and, possibly (in the future) new research findings based on the shared Galaxy Zoo data.

## V. Analysis and Implications

In the introduction I highlighted several results from the application of the knowledge commons research framework to Galaxy Zoo. Here I elaborate on and extend those results. I divide them into analysis and implications with respect to commons governance, first, and then with respect to the research framework itself.

First, at a high conceptual level, Galaxy Zoo confirms the existence of a dynamic relationship among scientific practice, forms of knowledge and knowledge structures, and social organization. Galaxy Zoo also suggests a similar dynamic relationship with respect to commons governance. Scientific practice has no fixed or necessary institutional or organizational form. Its forms are adapted dynamically to the interests of relevant scientific communities and to the characteristics of the research problems to be addressed and the data to be acquired and tested. The four attributes of Mertonian science are present in this case, but they are put to the test in some respects by the organizers of Galaxy Zoo aggressively building Galaxy Zoo with entrepreneurial zeal and tactics. One attribute of successful entrepreneurs is the ability to "pivot" from one project to the next, whether turning from failure to new opportunity or capitalizing on and institutionalizing early success. The Zooniverse, as an extension of Galaxy Zoo beyond astronomy, seems to be a case of the latter. In the process of Galaxy Zoo's evolution, its commons character evolved as well. What began as community construction to solve questions about galaxies evolved over time into the development of an infrastructural resource—a citizen-science peer-production tool.

Second, at a more concrete level, specific forms of social organization—including both the shape of astronomy and astrophysics disciplines and the character of their commons governance—are dependent on elastic conceptual and material (technologically grounded) understandings of the data that scientists generate and use. Modern astronomy and astrophysics are large and complex exercises in information and knowledge governance in addition to celestial observation. The computational character of the disciplines is in many respects akin to the computational character of contemporary geophysics. In light of that elasticity, none of these disciplines is itself static. What "is" science (or what "is" astronomy), what "is" relevant scientific research, and how individuals and groups bind themselves to those understandings, tightly or loosely, are parts of the processes by which commons is constituted, and by which commons constitutes those disciplines. I suggest, in other words, generalizing Kelty's observation about "recursive publics" in the context of open source software projects. Galaxy Zoo not only produces knowledge; it is in effect a form of knowledge itself. Governance (in this instance, commons) and knowledge and are mutually and recursively constitutive. One exists largely because of the other, and they evolve together.

Finally, and despite the popular view that citizen science projects such as Galaxy Zoo primarily involve participation by undifferentiated "crowds" of volunteers, the knowledge community that makes up Galaxy Zoo in this case succeeds in large part because it incorporates a well-defined (though far from rigid) social structure, with leaders, managers, and followers, and a layered matrix of visions and purposes, from "this is science" to "this is astronomy" to "this is how the astronomical data should be classified," that both produce that social structure and are reinforced by it. In part this is a rhetorical strategy that has the effect of distinguishing the work of volunteer scientists from that of unpaid laborers. In part it seems to be part of a larger, implicit strategy to align the volunteers with the professionals as part of the scientific enterprise. Throughout this chapter I have noted the relevance of appeals to the vision of Galaxy Zoo and to the scientific narrative as constituting a meaningful part of the project's success. The implication is this: If everyone is on the same page from the beginning, then it may be far less necessary to develop and use formal discipline to keep everyone in line with collective expectations and goals.

How do people get on the same page to begin with, and how do we know whether the vision continues to bind them together? Galaxy Zoo benefited from the existence of a large preexisting pool of amateur astronomers and the availability of an enormous, publicly accessible pool of observational data. Galaxy Zoo also benefited from the fact it was able to minimize the gap between what volunteers were doing (actually examining observational data) and what the professionals would do (in fact, the very same thing), despite its reliance on very large-scale and very complex information technology infrastructures. By expanding the opportunities for participation in "real" science, via the Galaxy Zoo forum, and by nominally including volunteers in Galaxy Zoo papers, the Galaxy Zoo team cultivated a meaningful scientific identity among the volunteers. The fact that the volunteers produced high-quality work, and at times even scientifically innovative work, created a virtuous circle, ratifying the initial decision to undertake Galaxy Zoo, reinforcing the collaborative character of Galaxy Zoo and providing a template for extending Galaxy Zoo to other research questions. One should always been alert for signs of exploitation in a case such as this, but there appear to be very few of them here. Galaxy Zoo volunteers were not trained expert researchers, but the work that they were doing was real and important enough that it is not implausible to treat it under the big tent of science. This finding is consistent with the research of the sociologist Harrison White (2008), which models persistent social organizations in terms of patterns of relations, including identity, rather than in terms of the attributes and attitudes of individuals.

The suggestions that the social structure of commons matters a great deal in understanding the effectiveness of commons, and that vision and personal and group identity matter a great deal as well—more so, perhaps that individual knowledge, attributes, and preferences—suggest some modest refinements to the knowledge commons research framework and some hypotheses for future inquiry That framework, as we have seen, focuses much of its attention on the resources that are created and shared within a commons regime and on how individuals make decisions regarding those resources. "Tragedy of the commons" overconsumption dilemmas are typical starting points for investigating cases of commons governance. But whatever the problems or dilemmas may be that give rise to governance challenges, it is typical to frame analysis in terms of decisions by individual actors or agents with respect to the character of specified resources.

More prominent than individual decisions regarding resources and more influential in the success of Galaxy Zoo, I believe, are the presence of a defined research problem to be solved, the need to choose and/or to design an organizational form well suited to solving it, and efforts to define both of those things publicly in terms of a specific vision that linked peer-production to citizen science to data-driven scientific discovery. A kind of collective identity bound members of the group via rhetorical interventions and related sociotechnical designs. Future work on knowledge commons should address from the beginning the underlying problem being solved, which may or may not depend directly on legally determined attributes of the resources in question; the choice of institutional or organizational form with respect to how that problem is being solved; and the ways in which that choice of form leads to or is connected to modes of engagement by individuals.

## VI. Conclusion

This chapter has examined Galaxy Zoo, a scientific research enterprise that exists at the intersection of three phenomena: citizen science, peer production, and data-intensive science. Galaxy Zoo was founded in 2007 by astronomers at the University of Oxford as a way to enlist public volunteers to assist with data classification in connection with understanding the evolution of galaxies. Based on the number of volunteers who participated, the amount of data processed, the speed and accuracy with which the project was completed, and the number of scholarly research papers produced, Galaxy Zoo has been a notable success. Its organizers have applied the same crowdsourcing techniques to large-scale data analysis in follow-on astronomy projects and in other areas of scientific research.

Here, Galaxy Zoo is examined not for its own sake but as a case of commons governance, using the knowledge commons research framework proposed by Madison, Frischmann, & Strandburg (2010) and described in Chapter 1 of this volume. The framework was proposed as a way of systematically investigating knowledge governance regimes that rely on resource sharing in settings involving nondepletable resources such as knowledge and information. This case study determined that the most significant aspects of Galaxy Zoo, and the most important reasons for its effectiveness as commons, have less to do with the character of its information resources (scientific data) and rules regarding their usage, and more to do with the expanded community constructed from hundreds of thousands of Galaxy Zoo volunteers. That community was guided from the outset by a vision of a specific organizational solution to a specific research problem in astronomy, initiated and governed, over time, by professional astronomers in collaboration with their expanding universe of volunteers. Future knowledge commons research should be especially attentive to the social organization of commons as a key factor in their success and effectiveness, along with the dynamics of resource production and consumption.

## References

Tim Adams, *Galaxy Zoo and the New Dawn of Citizen Science*, The Guardian, Mar. 17, 2012, http://www.guardian.co.uk/science/2012/mar/18/galaxy-zoo-crowdsourcing-citizen-scientists.

George A. Akerlof & Rachel E. Kranton, Identity Economics: How Our Identities Shape Our Work, Wages, and Well-Being (Princeton University Press 2010).

Yochai Benkler, The Wealth of Networks: How Social Production Transforms Markets and Freedom (Yale University Press 2006).

Margaret L. Berendsen, *Conceptual Astronomy Knowledge among Amateur Astronomers*, 4 Astronomy Education Rev. 1 (2005).

Mario Biagioli, *Documents of Documents: Scientists' Names and Scientific Claims*, in Documents: Artifacts of Modern Knowledge 129 (Annelise Riles ed., University of Michigan Press 2006).

John Seely Brown & Paul Duguid, The Social Life of Information (Harvard Business Review Press 2000).

Janet Browne, Charles Darwin: A Biography, Vol. 1 – Voyaging (Princeton University Press 1996).

Adrian Burton, *The Ichthyosaur in the Room*, 10 Frontiers in Ecology and the Envir. 340 (2012).

Carie Cardamone, *The Story of the Peas: Writing a Scientific Paper*, Galaxy Zoo, A Zooniverse Project Blog (July 2, 2009), http://blog.galaxyzoo.org/2009/07/02/the-story-of-the-peas-writing-a-scientific-paper/.

Carolin N. Cardamone et al., *Galaxy Zoo Green Peas: Discovery of a Class of Compact Extremely Star-Forming Galaxies*, 399 Monthly Notices of the Royal Astronomical Soc'y 1191 (2009).

Daniel Clery, *Galaxy Zoo Volunteers Share Pain and Glory of Research*, 333 Science 173 (2011).

Julie E. Cohen, Configuring the Networked Self: Law, Code, and the Play of Everyday Practice (Yale University Press 2012).

August Comte, Course of Positive Philosophy (Hermann 1975) (originally published 1830) (translated and condensed as H. Martineau, The Positive Philosophy of August Comte (2 vols. London 1875) (1st ed. London 1853)).

Caren Cooper, *Victorian-Era Citizen Science: Reports of Its Death Have Been Greatly Exaggerated, Scientific American Blogs* (Aug. 30, 2012a), http://blogs.scientificamerican.com/guest-blog/2012/08/30/victorian-era-citizen-science-reports-of-its-death-have-been-greatly-exaggerated/.

Caren Cooper, *Retro Science, Part 1*, Scientific American Blogs (Aug. 23, 2012b), http://blogs.scientificamerican.com/guest-blog/2012/08/23/retro-science-part-1/.

Rebecca S. Eisenberg, *Patents and the Progress of Science: Exclusive Rights and Experimental Use*, 56 U. Chicago L. Rev. 1017 (1989).

Timothy Ferris, Seeing in the Dark: How Amateur Astronomers Are Discovering the Wonders of the Universe (Simon & Schuster 2002).

Ann K. Finkbeiner, A Grand and Bold Thing: An Extraordinary New Map of the Universe Ushering in a New Era of Discovery (Free Press 2010).

Lucy Fortson et al., *Galaxy Zoo: Morphological Classification and Citizen Science*, in Advances in Machine Learning and Data Mining for Astronomy 213 (Michael J. Way et al. eds., 2009).

Chiara Franzoni & Henry Sauermann, *Crowd Science: The Organization of Scientific Research in Open Collaborative Projects*, 43 Research Policy 1 (2014).

Ira H. Fuchs, *Prospects and Possibilities of the Digital Age*, 145 Proceedings of the Am. Philosophical Society No. 1 (Mar. 2001).

Alyssa A. Goodman & Curtis G. Wong, *Bringing the Night Sky Closer: Discoveries in the Data Deluge*, in The Fourth Paradigm: Data-Intensive Scientific Discovery 39 (Tony Hey et al. eds., Microsoft Research 2009).

Warren O. Hagstrom, The Scientific Community (Basic Books 1965).

Tony Hey, *Data Esperanto*, Tony Hey on eScience (Apr. 28, 2011), http://tonyhey.net/2011/04/28/data-esperanto/.

Tony Hey, *A Journey to Open Access—Part 3*, Tony Hey on eScience (Jan. 18, 2013), http://tonyhey.net/2013/01/18/a-journey-to-open-access-part-3/.

Tony Hey et al. (Eds.), The Fourth Paradigm: Data-Intensive Scientific Discovery (Microsoft Research 2009).

Stephen Hilgartner & Sherry Brandt-Rauf, *Data Access, Ownership, and Control: Toward Empirical Studies of Access Practices*, 15 Knowledge 355 (1994).

John Hules, *Discovery of Dark Energy Ushered in a New Era in Computational Cosmology*, NERSC (Oct. 4, 2011), https://www.nersc.gov/news-publications/news/nersc-center-news/2011/nobel-laureate-blazed-new-trails-in-computational-cosmology/discovery-of-dark-energy-ushered-in-a-new-era-in-computational-cosmology/.

Christopher M. Kelty, Two Bits: The Cultural Significance of Free Software and the Internet (Duke University Press 2008).

Thomas S. Kuhn, The Copernican Revolution: Planetary Astronomy in the Development of Western Thought (Harvard University Press 1957).

Thomas S. Kuhn, The Structure of Scientific Revolutions (University of Chicago Press 1962).

John Lankford, *Amateurs and Astrophysics: A Neglected Aspect in the Development of a Scientific Specialty*, 11 Social Studies of Science 275 (1981).

A. Lawrence, *Drowning in Data: VO to the Rescue*, in Astronomy: Networked Astronomy and the New Media (R. J. Simpson & D. Ward-Thompson eds., Canopus Publishing 2009).

Chris Lintott, *A Week Inside the Galaxy Zoo*, Chris Lintott's Universe (July 18, 2007), http://chrislintott.net/2007/07/18/a-week-inside-the-galaxy-zoo/.

Chris Lintott, Interviewed by Michael Madison, June 25, 2013.

Chris J. Lintott et al., *Galaxy Zoo: Morphologies Derived from Visual Inspection of Galaxies from the Sloan Digital Sky Survey*, 389 Monthly Notices of the Royal Astronomical Soc'y 1179 (2008).

Chris Lintott et al., *Galaxy Zoo 1: Data Release of Morphological Classifications for Nearly 900,000 Galaxies*, 410 Monthly Notices of the Royal Astronomical Soc'y 166 (2010).

Michael J. Madison, *Reconstructing the Software License*, 35 Loyola U. Chi. L.J. 275 (2003).

Michael J. Madison, *Law as Design: Objects, Concepts, and Digital Things*, 56 Case Western Reserve. L. Rev. 381 (2005).

Michael J. Madison, Brett M. Frischmann, & Katherine J. Strandburg, *Constructing Commons in the Cultural Environment*, 95 Cornell L. Rev. 657 (2010).

Thomas Mandeville, Understanding Novelty: Information, Technological Change, and the Patent System (Ablex 1996).

Robin Mansell, *Employing Digital Crowdsourced Information Resources: Managing the Emerging Information Commons*, 7 Int'l J. of the Commons 255 (2013).

James Manyika et al., Big Data: The Next Frontier for Innovation, Competition, and Productivity (McKinsey Global Institute 2011).

Bruce Margony, *The Sloan Digital Sky Survey*, 357 Philosophical Transactions of the Royal Society London A, No. 1750 93 (Jan. 15, 1999).

Karen L. Masters, *She's an Astronomer: Meg Urry*, Galaxy Zoo, A Zooniverse Project Blog (May 2, 2010), http://blog.galaxyzoo.org/2010/05/02/shes-an-astronomer-meg-urry/.

Karen L. Masters, *She's an Astronomer: Did We Really Need that Series?*, Galaxy Zoo, A Zooniverse Project Blog (Jan. 7, 2011), http://blog.galaxyzoo.org/2011/01/07/shes-an-astronomer-did-we-really-need-that-series/.

Karen L. Masters, *Invited Discourse: "A Zoo of Galaxies,"* 16 Highlights of Astronomy (XXVIIth IAU General Assembly, August 2009, Thierry Montmerle ed.) (2013), http://arxiv.org/pdf/1303.7118v1.pdf.

Christine McGourty, *Scientists Seek Galaxy Hunt Help*, BBC News (July 11, 2007), http://news.bbc.co.uk/2/hi/science/nature/6289474.stm.

Robert P. Merges, *Property Rights Theory and the Commons: The Case of Scientific Research*, 13 Social Philosophy & Policy, No. 2, at 145 (1996).

Robert K. Merton, *The Normative Structure of Science*, reprinted in The Sociology of Science: Theoretical Empirical Investigations 267 (Norman W. Storer ed., University of Chicago Press 1973) (originally published 1942).

National Research Council, The Future of Scientific Knowledge Discovery in Open Networked Environments (National Academy Press 2012).

Nature, *Special issue on "Big Data,"* Vol. 455, Issue no. 7209 (Sept. 4, 2008).

Sue Nelson, *Big Data: The Harvard Computers*, 455 Nature 36 (Sept. 4, 2008).

Sarah Ogilvie, Words of the World: A Global History of the Oxford English Dictionary (Cambridge University Press 2012).

Elinor Ostrom, Governing the Commons: The Evolution of Institutions for Collective Action (Cambridge University Press 1990).

Elinor Ostrom, *Background on the Institutional Analysis and Development Framework*, 39 The Policy Studies J. 7 (2011).

Michael Polanyi, Science, Faith and Society (Oxford University Press 1946).

M. Jordan Raddick et al., *Galaxy Zoo: Exploring the Motivations of Citizen Science Volunteers*, 9 Astronomy Education Rev. 010103-1 (2010).

Arti Kaur Rai, *Regulating Scientific Research: Intellectual Property Rights and the Norms of Science*, 94 Northwestern U. L. Rev. 77 (1999).

Jason Reed et al., *A Framework for Defining and Describing Key Design Features of Virtual Citizen Science Projects*, in Proceedings of the 2012 iConference 623 (Toronto 2012).

J. H. Reichman & Paul F. Uhlir, *A Contractually Reconstructed Research Commons for Scientific Data in a Highly Protectionist Intellectual Property Environment*, 66 Law & Contemp. Probs. 315 (2003).

Kevin Schawinski, Interviewed by Michael Madison, Dec. 1, 2011.

Kevin Schawinski, et al., *Galaxy Zoo: A Sample of Blue Early-type Galaxies at Low Redshift*, 396 Monthly Notices of the Royal Astronomical Soc'y 818 (2009).

Science, *Special issue on "Dealing with Data,"* Vol. 331, Issue no. 6018 (Feb. 11, 2011).

Alice [Sheppard], *Peas in the Universe, Goodwill and a History of Zooite Collaboration on the Peas Project*, Galaxy Zoo, A Zooniverse Project Blog (July 7, 2009), http://blog.galaxyzoo.org/2009/07/07/peas-in-the-universe-goodwill-and-a-history-of-zooite-collaboration-on-the-peas-project/.

Petr Škoda, *The Virtual Observatory and Its Benefits for Amateur Astronomers*, 75 Open European J. on Variable Stars 32 (2007).

Alexander S. Szalay et al., *The SDSS SkyServer—Public Access to the Sloan Digital Sky Server Data*, SIGMOD '02: Proceedings of the 2002 ACM SIGMOD International Conference on Management of Data (2002).

Sharon Traweek, Beamtimes and Lifetimes: The World of High Energy Physicists (Harvard University Press 1988)

William Whewell, The Philosophy of the Inductive Sciences: Founded upon Their History, Vol. 1 (London 1840).

Harrison C. White, Identity and Control: How Social Formations Emerge (2d ed. Princeton University Press 2008)

Simon Winchester, The Meaning of Everything: The Story of the Oxford English Dictionary (Oxford University Press 2004).

Donald G. York et al., *The Sloan Digital Sky Survey: Technical Summary*, 120 The Astronomical J. 1579 (2000).